%% file: Cufaro_Sabino_Fast_Pricing.tex
\documentclass[showpacs,preprintnumbers,amsmath,amssymb,12pt]{article}

\usepackage{a4wide,amsmath,amsthm,epsfig,graphicx}
\usepackage{amsmath,amsthm,amssymb,eurosym}
\usepackage{amsfonts}
\usepackage{graphics}
\usepackage{adjustbox}

\usepackage{dsfont}
\usepackage{bbm}
\usepackage{bm}
\usepackage{xcolor}
\usepackage{dcolumn}
\usepackage{fancyhdr,fancybox}
\usepackage[svgnames]{pstricks}
\usepackage{pst-text}
\usepackage{pst-plot,pst-eucl,pstricks-add}
\usepackage{caption}
\usepackage{subcaption}

\usepackage{xpatch}
\usepackage{algpseudocode}

\usepackage{algorithm}
\usepackage{algpseudocode}
\newcommand{\refeq}[1]{~(\ref{#1})}
\newcommand{\myref}[1]{~\ref{#1}}
\newcommand{\mycite}[1]{~\cite{#1}}

\newcommand{\rv}{\textit{rv}}
\newcommand{\id}{\textit{id}}
\newcommand{\iid}{\textit{iid}}
\newcommand{\sd}{\textit{sd}}

\newcommand{\pdf}{\textit{pdf}}

\newcommand{\chf}{\textit{chf}}

\newcommand{\Pqo}{\bm{P}\hbox{-\emph{a.s.}}}
\newcommand{\eqd}{\stackrel{d}{=}}

\newcommand{\PR}[1]{\bm{P}\left\{{#1}\right\}}
\newcommand{\EXP}[1]{\bm{E}\left[{#1}\right]}

\newcommand{\arem}{$a$-remainder}

\newcommand{\poiss}{\mathfrak{P}}
\newcommand{\pol}{\overline{\mathfrak{B}}}

\newcommand{\unif}{\mathfrak{U}}

\newcommand{\gam}{\Gamma}
\newcommand{\bgam}{\mathfrak{b}\Gamma}

\newcommand{\erl}{\mathfrak{E}}

\newcommand{\bin}{\mathfrak{B}}

\newcommand{\ou}{OU}
\newcommand{\BM}{\emph{BM}}

\newcommand{\Levy}{L\'{e}vy}
\newcommand{\gou}{$\Gamma$-OU}
\newcommand{\bgou}{$\mathrm{bi}\Gamma$-OU}

\title{\LARGE \textbf{ Fast Pricing of Energy Derivatives\\ with Mean-reverting Jump-diffusion Processes}\footnote{ The views, opinions, positions or strategies
expressed in this article are those of the authors and do not
necessarily represent the views, opinions, positions or strategies
of, and should not be attributed to E.ON SE.}}

\author{Nicola \textsc{Cufaro Petroni}\footnote{cufaro@ba.infn.it}  \\
Dipartimento di \textsl{Matematica} and \textsl{TIRES}, Universit\`a di Bari\\
\textsl{INFN} Sezione di Bari\\ \vspace{7pt}
via E. Orabona 4, 70125 Bari, Italy\\
Piergiacomo
\textsc{Sabino}\footnote{piergiacomo.sabino@eon.com}\\
Quantitative Modelling \\
E.ON SE\\
\vspace{5pt}
 Br\"{u}sseler Platz 1, 45131 Essen, Germany
}

\date{}
\begin{document}
    \maketitle \thispagestyle{empty}
        \begin{abstract}
				Most energy and commodity markets exhibit mean-reversion and occasional distinctive price spikes,
which results in demand for derivative products which protect the holder against
high prices.

To this end, in this paper we present exact and fast methodologies for the simulation of the spot price dynamics modeled as the exponential of the sum of an Ornstein-
Uhlenbeck and an independent pure jump process, where the latter one is driven by a compound Poisson process with (bilateral) exponentially distributed jumps.
These methodologies are finally applied to the pricing of Asian options, gas storages and swings under different combinations of jump-diffusion market
models, and the apparent computational advantages of the proposed
procedures are emphasized.
        \end{abstract}

        \section{Introduction and Motivation}
The mathematical modeling of the day-ahead price in commodity and
energy markets  is supposed to capture some peculiarities like
mean-reversion, seasonality and jumps.  A typical approach consists
in resorting to price processes driven either by a generalized
Ornstein-Uhlenbeck (\ou) process, or by a regime switching process.
The  current literature is very rich of model  suggestions: Lucia
and Schwartz\mycite{LS02}, for instance, propose a one-factor
Gaussian-\ou\ with application to the Nordic Power Exchange, whereas a
two factor version can be found in Schwartz and Smith\mycite{SchwSchm00} with an additional Brownian Motion (\BM). Models
that go beyond the Gaussian world can be found among others in
Benth et al.\cite{BMBK07}, Meyer-Brandis  and Tankov\mycite{MBT2008} and Cartea and Figueroa\mycite{CarteaFigueroa}. The
first two papers investigate the use of generalized \ou\ processes,
while the last one studies the modeling with a jump-diffusion \ou\
process and a regime switching. In the present paper we  first analyze the properties of a
mean-reverting \ou\ process driven by a compound Poisson process
with exponential jumps  superposed to a standard Gaussian \ou\
process. This  combination has been investigated also by other
authors: for instance Deng\mycite{Deng00stochasticmodels}, Kluge\mycite{Kluge2006} and Kjaer\mycite{Kjaer2008}, or  even Benth and
Pircalabu\mycite{BenthPircalabu18} in the context of modeling wind
power futures.

Having selected a market model driven by a mean-reverting jump-diffusion dynamics, it is quite common to use Monte Carlo methods to price derivative contracts.
To this end, it is quite important to design fast and efficient simulation procedures particularly for real-time pricing. Indeed, risk management and trading units have to deal with a large number of contracts whose prices and sensitivities have to be evaluated regularly and, of course, the computational time may become an issue.
The simulation of the skeleton of a Gaussian-driven \ou\ process is standard and efficient, whereas the generation of the path of a \ou\ process with exponential jumps deserves particular attention. The simulation of this latter process can be based on the process definition itself, for example using a modified version of the Algorithm 6.2 page 174 in Cont and Tankov\mycite{ContTankov2004}.
Although sometime referred with different naming convention, a mean-reverting compound Poisson process with exponential jumps is known in the literature as Gamma-\ou\ process (\gou) because it can be proven that its marginal law is a gamma law (see Barndorff-Nielsen and Shephard\mycite{BNSh01}).

Recently, two different approaches have been proposed to address the simulation of a \gou\ process. Based on the decomposition of the \ou\ process into simple components, Qu et al.\mycite{QDZ19} propose an exact simulation procedure that has the advantage of avoiding the simulation of the jump times. On the other hand, in Cufaro Petroni and Sabino\mycite{cs20} we have studied the distributional properties of a \gou\ and bilateral-\gou\ process (\bgou) and found the density and characteristic function in close form. In particular, we have proven that such a law can be seen as a mixture of well-known laws giving, as a side-product, very fast and efficient simulation algorithms.

In this work we compare the computational performance of the new and traditional algorithms in the context of pricing complex energy derivatives, namely Asian options, gas storages and swings, that normally require a high computational effort. We consider three types of market models via the superposition of a Gaussian-driven \ou\ process to three different combination of \gou\ and \bgou\ processes.
The numerical experiments that we have conducted show that our algorithms outperform any other approaches and can provide a remarkable advantage in terms of computational time which constitute the main contribution of this paper. In the worst case, it is thirty times faster for the pricing of Asian options and ``only''  forty percent faster for storages and swings using a Monte Carlo based stochastic optimization. Our results demonstrate that our methodology is by far the best performing and is suitable for real-time pricing.

The paper is structured as follows: in
Section\myref{sec:markets} we introduce the three market models driven by a mean-reverting jump-diffusion dynamics that we will adopt for the pricing of the energy derivatives.
Section\myref{sec:gen:ou}
introduces the concept of generalized \ou\
processes and the details the algorithms available for the exact
simulation of a \gou\ or a \bgou\ process.
Section\myref{sect:numExperiments} illustrates the extensive numerical experiments that we have conducted. As mentioned, we consider the pricing of Asian options, gas storages and swings.
Finally, Section\myref{sec:conclusions}
concludes the paper with an overview of future inquiries and
possible further applications.

        \section{Market Models \label{sec:markets}}

From the financial perspective, it is well-known that
day-ahead prices exhibit seasonality, mean reversion and jumps, therefore
a realistic market model has to  capture these features.

Similarly to Kluge\mycite{Kluge2006} and Kjaer\mycite{Kjaer2008}, in this study, we assume that the dynamics of the day-ahead (spot) price can be decomposed into three independent factors
            \begin{eqnarray}
                S(t) &=& F(0,t)\, \exp\left\{h(t) + \sum_{d=1}^{D}X_d(t) + \sum_{j=1}^{J}Y_j(t)\right\} =  F(0,t)\, e^{h(t) + H(t)}\nonumber\\
                 &=& S_{\mathrm{season}}(t) \,{\cdot}  S_{\mathrm{diffusion}}(t) \,{\cdot}  S_{\mathrm{jumps}}(t),
            \end{eqnarray}
                        where denoting $\varphi_H(u,t) = \EXP{iu\,H(t)}$, $\varphi_{X_d(t)}(u,t) = \EXP{iu\,X_d(t)}$ and $\varphi_{Y_j(t)}(u,t) = \EXP{iu\,Y_j(t)}$ we have
            \begin{equation}\label{eq:chf_spot}
                \varphi_H(u,t) =\prod_{d=1}^{D}\varphi_{X_d}(u,t)\prod_{j=1}^{J}\varphi_{Y_j}(u,t)= \varphi_{\mathrm{diffusion}}(u,t) \,
                 {\cdot}\,\varphi_{\mathrm{jumps}}(u,t)
            \end{equation}
Using the risk-neutral arguments  of the Lemma 3.1 in Hambly
et al.\mycite{HHM11}, we get the deterministic function $h(t)$
consistent with forward curve
            \begin{equation}\label{eq:rn:spot}
              h(t) = -\log\varphi_H(-i, t).
            \end{equation}
In particular, we consider the following representation of spot prices
\begin{equation}\label{eq:spot}
              S(t) = F(0,t)\, e^{h(t) + X(t) + Y(t)}
\end{equation}
with only one standard Gaussian OU process
\begin{eqnarray}
X(t) &=& X(0)e^{-\varrho\,t} + \sigma\int_{0}^{t}e^{-\varrho(t-s)}dW(s)\\
\log\varphi_{\mathrm{diffusion}}(u,t)&=&  iuX(0)e^{-\varrho\,t} -\frac{u^2\sigma^2}{4\varrho}\left(1 - e^{-2\varrho \,t}\right).
\end{eqnarray}
We do not consider any additional BM as done in Schwartz and Smith
\cite{SchwSchm00}, but we assume that $Y(t)$ follows one of the three dynamics below.
\begin{enumerate}
	\item[Case 1]
	\begin{equation}\label{eq:spot:jump:kou}
             Y(t) = Y(0)e^{-k\,t} + \sum_{n=1}^{N(t)}e^{-k(t-\tau_n)}J_n\\
            \end{equation}
where $N(t)$ is a Poisson process with intensity $\lambda$ and jump times $\tau_n$; $J_n$ are then distributed according to a double exponential distribution as defined in Kou\mycite{Kou2002}, namely a mixture of a positive exponential \rv\ $U\sim\erl_1(\beta_1)$ and a negative exponential \rv\ $-D\sim\erl_1(\beta_2)$ having mixture parameters $p$ and $q=1-p$ with the following \pdf\ and \chf
\begin{equation}
	f_{\beta_1,\beta_2, p}(x)=p\beta_1e^{-\beta_1 x}\mathds{1}_{x\ge 0} + (1-p)\beta_2 e^{\beta_2 x}\mathds{1}_{x< 0}
\label{eq:double:exp:pdf}
\end{equation}
\begin{equation}
\varphi_{\beta_1,\beta_2, p}(v)  =
  p\frac{\beta_1}{\beta_1-iv} + (1-p) \frac{\beta_2}{\beta_2+iv} = p\varphi_u(v) + (1-p)\varphi_d(v).
\end{equation}
It means that each \rv\ $J_n$ can be seen as $J_n\eqd B_n\,U_n - (1-B_n)D_n$ where $B_n$ is a binomial \rv\ with distribution $\bin(p)$.
Without loss of generality let $Y(0)=0$, as shown in Cufaro Petroni and Sabino\mycite{cs20},  the jump process $Y(t)=\sum_{n=1}^{N(t)}e^{-k(t-\tau_n)}J_n$ can be seen as the difference of two independent processes $Y(t) =Y_1(t) - Y_2(t)$ with $Y_1(t)=\sum_{n=1}^{N_1(t)}e^{-k(t-\tau_n)}U_n$ and $Y_2(t)=\sum_{n=1}^{N_2(t)}e^{-k(t-\tau_n)}D_n$, with the same parameter $k$,
where now $N_1(t)$ and $N_2(t)$ are two independent Poisson processes with intensities $\lambda_1=p\lambda$ and $\lambda_2=(1-p)\lambda$, respectively. Hence,
\begin{equation}
\varphi_{\mathrm{jumps}}(u,t) = \varphi_1(u, t) \varphi_2(-u, t)
\label{eq:chf:mm1}
\end{equation}
where  $\varphi_1(u, t)$ and $\varphi_2(u, t)$ are the \chf's of a mean-reverting Poisson process with (upward) exponentially distributed jumps at time $t$ with rates $\beta_1$ and $\beta_2$, respectively.
	\item[Case 2]
	\begin{eqnarray}\label{eq:spot:jump:asymmetric}
		Y(t)&=& Y_1(t) - Y_2(t)\\
		Y_1(t) &=& Y_1(0)e^{-k_1\,t} + \sum_{n=1}^{N_1(t)}e^{-k_1(t-\tau_n^{(1)})}U_n\nonumber\\
		Y_2(t) &=& Y_2(0)e^{-k_2\,t} + \sum_{m=1}^{N_2(t)}e^{-k_2(t-\tau_m^{(2)})}D_m,
\end{eqnarray}
\noindent where $N_1(t)$ and $N_2(t)$ are two independent Poisson processes with intensities $\lambda_1$ and $\lambda_2$, respectively and
$U_n$ and $U_m$ are independent  \rv's with exponential laws $\erl_1(\beta_1)$ and $\erl_1(\beta_2)$ respectively.
	\item[Case 3]
	The jumps $J_n$ of the process $Y(t) = Y(0)e^{-k\,t} + \sum_{n=1}^{N(t)}e^{-k_N(t-\tau_n)}J_n$ are now distributed according to a centered Laplace  \rv's with parameter $\beta$. This jump process can also be seen as the difference of two independent processes $Y(t) =Y_1(t) - Y_2(t)$ as in\refeq{eq:spot:jump:asymmetric}, where here $Y_1(t)$ and $Y_2(t)$ have the same parameter $k$ and  $U_n$ and $D_m$ independent \rv's with the same laws $\erl_1(\beta)$.
\end{enumerate}

The simulation of a Gaussian-driven OU process is standard and very fast whereas on the other hand, the building block for the simulation of each of the jump processes introduced above is the generation of a \rv\ distributed according to the law of a compound Poisson process with exponential jumps. Therefore, the overall computational effort will be deeply affected by that required to simulate the jump process.
To this end, the simulation procedure of the skeleton of the
day-ahead price $S(t)$ in\refeq{eq:spot} over a time grid $t_0, t_1,\dots, t_M$
($\Delta t_m = t_m - t_{m-1}\,,\; m=1,\dots,M $) consists in the steps illustrated in Algorithm\myref{alg:spot}.
\begin{algorithm}
\caption{ }\label{alg:spot}
		\begin{algorithmic}[1]
		\For{ $m=1, \dots, M$}
		\State $h(t_m)\gets -\frac{\sigma^2}{4k}\left(1 - e^{-2k_D \,\Delta t_m}\right) - \log\varphi_{\mathrm{jumps}}(u,t_m)$
		\State Generate $x\sim\mathcal{N} \left(0, \sigma \sqrt{\frac{1 - e^{- 2k_D \Delta t_m}}{2k_D}}\right)$
		\State Generate $y_1\eqd \sum_{n=1}^{N_1(t_m)}e^{-k_1(t_m-\tau_n^{(1)})}U_n$
		\State Generate $y_2\eqd \sum_{\ell=1}^{N_2(t_m)}e^{-k_2(t_m-\tau_{\ell}^{(2)})}D_n$
		\State $X(t_m)\gets X(t_{m-1})e^{-k_D\Delta t_m} + x$
		\State $Y_i(t_m)\gets Y_i(t_{m-1})e^{-k_i\Delta t_m} + y_i$, $i=1,2$.
		\State $S(t_m)\gets e^{h(t_m) + X(t_m) + Y_1(t_m) - Y_2(t_m)}$
		\EndFor
		\end{algorithmic}
\end{algorithm}

Although sometimes the jump process with exponential jumps is mentioned under different names in the financial literature (e.g. \emph{MRJD }in Kjaer\mycite{Kjaer2008}),  such a process is known as Gamma-OU process (\gou), because its stationary law is a gamma distribution. In addition, being $Y(t)$ the difference of two \gou\ processes, one can show  that it coincides with a bilateral-gamma-OU process, denoted here \bgou\ (see Cufaro Petroni and Sabino\mycite{cs20} and K\"{u}chler and Tappe\mycite{KT2008}).

Finally, the exact simulation of the skeleton of $Y(t)$ depends on a fast generation of the \rv\ $y_i$ distributed according to the law of a \gou\ process at time $t$. We consider three alternative simulation algorithms available in the literature as discussed in the following section.

\section{Simulation of a \ou\ process with Compound Poisson noise\label{sec:gen:ou}}
Consider a \Levy\ process $Z(t)$, with $Z(1)$ distributed as
$\widetilde{\mathfrak{D}}$, and acting as the backward driving \Levy\ process (\emph{BDLP}) for the
generalized \ou-$\widetilde{\mathfrak{D}}$ process $Y(t)$ whose solution
is
            \begin{equation}\label{eq:genOU_solution}
              Y(t) = y_0e^{-kt} + \int_{0}^{t}e^{-k(t-s)}dZ(s).
            \end{equation}
Following Barndorff-Nielsen and Shephard
\cite{BNSh01}, given a distribution
$\mathfrak{D}$, we can find an infinitely divisible (\id) $\widetilde{\mathfrak{D}}$ such
that the \ou-$\widetilde{\mathfrak{D}}$ process $Y(t)$ is also
$\mathfrak{D}$-\ou\ (i.e.\ admits $\mathfrak{D}$ as stationary
distribution), if and only if $\mathfrak{D}$ is \emph{self-decomposable} (\sd).

We recall that a law with probability density (\pdf) $f(x)$ and
characteristic function (\chf) $\varphi(u)$ is said to be
 \sd\ (see Sato\mycite{Sato} or Cufaro
Petroni~\cite{cufaro08}) when for every $0<a<1$ we can find another
law with \pdf\ $g_a(x)$ and \chf\ $\chi_a(u)$ such that
                \begin{equation}\label{sdec}
                    \varphi(u)=\varphi(au)\chi_a(u)
                \end{equation}
We will accordingly say that a random variable (\rv) $X$ with \pdf\
$f(x)$ and \chf\ $\varphi(u)$ is \sd\ when its law is \sd: looking
at the definition this means that for every $0<a<1$ we can always
find two independent \rv's, $Y$ (with the same law of $X$)
and $Z_a$ (here called \emph{\arem}, with \pdf\ $g_a(x)$
and \chf\ $\chi_a(u)$) such that
                \begin{equation}\label{sdec-rv}
                    X\eqd aY+Z_a.
                \end{equation}
Consider now the process $Z(t)$
            \begin{equation*}
              Z(t)=\sum_{n=0}^{N(t)}J_n\qquad\quad
              J_0=0\quad\Pqo,
            \end{equation*}
with intensity $\lambda$ of
the number process $N(t)$, and identically distributed exponential
jumps $J_n\sim\erl_1(\beta)$ acting as  the \emph{BDLP} of the process $Y(t)$ in\refeq{eq:genOU_solution}. It is well-know (see for instance Schoutens\mycite{Schoutens03} page 68) that the stationary law of the latter process is a gamma distribution, therefore, such a process can be synthetically dubbed \gou($k, \lambda,\beta$) to recall its parameters.

Using this naming convention, the jump components of the three market models of Section\myref{sec:markets} are simply the difference of two \gou\ processes, also know as \bgou\ process, where in particular, the third market model is a symmetric \bgou\ process.

\subsection{Exponential jumps: \gou\ process \label{subsect:gou}}
A straightforward way to simulate the innovation $y$ of a \gou\ process with parameters $k, \lambda,\beta$ (used in the step four of Algorithm\myref{alg:spot}) simply consists in  adapting Algorithm 6.2 page 174 in Cont and Tankov\mycite{ContTankov2004} as detailed in Algorithm\myref{alg:ct}. It is 

\begin{algorithm}
\caption{ }\label{alg:ct}
\begin{algorithmic}[1]
		\State Generate $N\sim\poiss(\lambda\Delta t_m)$,\Comment {Poisson \rv~with intensity $\lambda\Delta t_m$}
		\State Generate $N$ \iid\ uniform \rv's $\bm{u}=(u_1, \dots, u_N)\,\sim\,\unif([0,1]^N)$.
		\State Sort $\bm{u}$, $(u_{[1]}< \dots< u_{[N]})$,
		\State $\tau_{n}\gets\Delta t_m u_{[{n}]},\, n=1,\dots N$,
		\State Generate $N$ \iid\ $J_{n}\sim\erl_1(\beta), n=1, \dots N$,  \Comment {Exponential \rv~ with rate $\beta$}
		\State $y\gets \sum_{{n}=1}^{N}e^{-k(\Delta t_m-\tau_{n})}J_{n}$.
		\end{algorithmic}
\end{algorithm}

Algorithm\myref{alg:ct} does not directly rely on the
statistical properties of the process $Y(t)$, but is rather based on its definition.

Starting from a different point of view, we have proposed in\mycite{cs20} two simulation algorithms that are fully based on the distributional properties of the \gou\ process.
One result shown in Cufaro Petroni and Sabino\mycite{cs20}
is that the law of a \gou\ process $Y(t)$ at time $t$ with parameters $(k, \lambda, \beta)$  coincides with that of the \arem\ $Z_a$ of a gamma law $\gam(\alpha,\beta)$ with scale parameter $\alpha=\lambda/k$ and rate parameter $\beta$ if one assumes $a=e^{-kt}$.

We recall that the laws of the gamma family $\gam(\alpha,\beta)$
($\alpha>0,\beta>0$) have the following \pdf\ and \chf
\begin{eqnarray}
  f_{\alpha,\beta}(x) &=& \frac{\beta}{\Gamma(\alpha)}(\beta x)^{\alpha-1}e^{-\beta x} \qquad\qquad x>0\label{gammapdf}\\
  \varphi_{\alpha,\beta}(u)  &=&
  \left(\frac{\beta}{\beta-iu}\right)^\alpha\label{gammachf}
\end{eqnarray}
\noindent In particular $\gam(k,\beta)$, with
$\alpha=k=1,2,\ldots$ a natural number, are the Erlang laws
$\erl_k(\beta)$, and $\gam(1,\beta)$  is the usual exponential
law $\erl_1(\beta)$.

Now consider a \rv\ $S$ distributed according to a
\emph{negative binomial, or Polya distribution}, denoted hereafter
$\pol(\alpha,p)\,,\; \alpha>0,\,0<p<1$,  namely such
that
\begin{equation*}
		\PR{S=k} =  \binom{\alpha+k-1}{k}(1-p)^\alpha p^k\qquad\quad k=0,1,\ldots
\end{equation*}
in Cufaro Petroni and Sabino\mycite{cs20} we have proven that the \pdf\ and \chf\ of $Z_a$ are
\begin{equation}
\chi_a(u, \alpha, \beta)=\left(\frac{\beta-iau}{\beta-iu}\right)^\alpha=\sum_{k=0}^\infty\binom{\alpha+k-1}{k}a^\alpha(1-a)^k\left(\frac{\beta}{\beta-iau}\right)^k
\label{eq:polya:gamma:chf}
\end{equation}
\begin{equation}\label{eq:mix:polya}
	g_a(x, \alpha, \beta)=a^\alpha\delta(x)+\sum_{k=1}^\infty\binom{\alpha+k-1}{k}a^\alpha(1-a)^kf_{k,\,\!^\beta/_a}(x) \qquad\qquad x>0
\end{equation}						
namely $Z_a$ is distributed according to the law of an infinite Polya
$\pol(\alpha,1-a)$-weighted mixture of Erlang laws
$\erl_k\big(\,\!^\beta/_a\big)$.
This distribution can also be
considered either as an Erlang law $\erl_S\big(\,\!^\beta/_a\big)$
with a Polya $\pol(\alpha,1-a)$-distributed random index $S$,  or
even as that of a sum of a Polya random number of \iid\ exponential \rv's.
\begin{equation*}
		\sum_{j=0}^SX_j\qquad\qquad
		S\sim\pol(\alpha,1-a)\qquad
		X_j\sim\erl_1\big(\,\!^\beta/_a\big)\qquad X_0=0,\;\Pqo
\end{equation*}
Based on the observations above, the \chf\ of a \gou\ process at time $t$ is
 \begin{equation}\label{eq:chf:gou}
\varphi(u, t)=\left(\frac{\beta-iue^{-kt}}{\beta-iu}\right)^{\frac{\lambda}{k}}
\end{equation}
and the simulation of the innovation $y$ of a \gou$(k, \lambda, \beta)$ process is then shown in Algorithm\myref{alg:cs}.
\begin{algorithm}
\caption{ }\label{alg:cs}
		\begin{algorithmic}[1]
		\State $\alpha\gets\lambda/k,\;\; a\gets e^{-k\Delta t_m}$
		\State $b\gets B\sim \pol(\alpha,1-a)$ \Comment{Generate a Polya $(\alpha,1-a)$ \rv}
		\State $y \gets \erl_{b}\left(\beta/a\right)$; \Comment{Generate an Erlang \rv\ with rate $\beta/a$}
		\end{algorithmic}
\end{algorithm}
It is worthwhile noticing that such an algorithm resembles to the one proposed in McKenzie \cite{McK87} with the advantage to simulate Erlang \rv's only.

A different methodology to simulate a \gou\ has been recently proposed in Qu et al.\mycite{QDZ19} and it is based on the following different representation of the conditional \chf\ of a \gou\ process
\begin{equation}
\EXP{e^{iuY(t+s)|Y(s)}}=e^{iuY(s)e^{-kt}}\times e^{\lambda t \left(\varphi_{\tilde{J}}(u)-1\right)}
\label{eq:chf:qdz}
\end{equation}
where
\begin{equation}
\varphi_{\tilde{J}}(u)=\int_0^1\frac{\beta e^{ktv}}{\beta e^{ktv} - iu}dv,
\label{eq:chf:exp:qdz}
\end{equation}
that coincides with \chf\ of a compound Poisson process with exponentially distributed jumps with random rate $\tilde{\beta}\eqd\beta e^{k\Delta t U}$, and $U\sim\,\unif([0,1])$.

This third procedure is summarized in Algorithm\myref{alg:qdz}.
\begin{algorithm}
\caption{ }\label{alg:qdz}
\begin{algorithmic}[1]
		\State Generate $N\sim\poiss(\lambda\Delta t_m)$,\Comment {Poisson \rv~with intensity $\lambda\Delta t_m$}
		\State Generate $N$ \iid\ uniform \rv's $\bm{u}=(u_1, \dots, u_N)\,\sim\,\unif([0,1]^N)$.
		\State $\tilde{\beta_n}\gets\beta e^{k\Delta t_m u_n}, n=1,\dots, N$.
		\State Generate $N$ \iid\ $\tilde{J}_n\sim\erl_1(\tilde{\beta_n}), n=1,\dots, N$, \Comment {Exponential \rv's with random rate $\tilde{\beta_i}$}
		\State $y\gets  \sum_{i=n}^{N}\tilde{J}_n$.
		\end{algorithmic}
\end{algorithm}

Algorithms\myref{alg:cs} and\myref{alg:qdz} avoid simulating the jump times of the Poisson process whereas on the other hand, Algorithm\myref{alg:ct} and Algorithm\myref{alg:qdz} require similar operations and  additional steps compared to Algorithm\myref{alg:cs} which, as observed in Cufaro Petroni and Sabino\mycite{cs20}, is by far the fastest alternative.

Finally, considering for simplicity an equally-spaced time
grid, one might be tempted (as often done) to use a Euler discretization
 with the assumption that only one jump
can occur within each time step with probability $\lambda\Delta
t$:
            \begin{equation}\label{eq:Euler}
              Y(t_m) = Y(t_{m-1})(1-k\Delta t) + B_m(1)J_m, \quad m=1,\dots M,
            \end{equation}
where $B_m(1)\sim\bin(1, \lambda\Delta t)$ are $m$
independent Bernoulli \rv's. Taking then for simplicity $b =
1-\lambda\Delta t$, the \chf\ of $B_m(1)Y_m$ is
            \begin{equation*}
              \varphi_m(u,t) = b + \beta\frac{1-b}{\beta - iu} = \frac{\beta - ibu}{\beta - iu}
               = \frac{\beta - i(1-\lambda \Delta t)u}{\beta - iu}
            \end{equation*}
This \chf\ however, could be considered as a first order
approximation of\refeq{eq:chf:gou} only if $k=\lambda$. Of course, a reduction of the time step would by no means provide an
improvement, and hence any calibration, or pricing of derivatives
relying on the simulation of an \gou\ with the assumption that
only one jump can occur per time step would lead to wrong and biased
results.

\subsection{ Time-dependent Poisson Intensity\label{subsect:tdLambda}}
Jumps are often concentrated in clusters, for instance energy
markets are very seasonal and jumps  more often occur
during  either  a period of high demand or  a period of cold
spell. A more realistic approach could then be to consider a
non-homogeneous Poisson process with time-dependent intensity
$\lambda(t)$ with $\Lambda(t)=\int_0^t\lambda(s)ds$.  In
this case, the new Poisson process and its relative compound version
have independent, but non-stationary increments. The
modeling then becomes more challenging
and somehow depends on the choice of the
specific intensity function. In any case, one could consider a time
grid $t_0, t_1,\dots, t_M$ fine enough such that the non-homogeneous Poisson process has a
step-wise intensity, $\lambda(s) = \lambda_m
\mathds{1}_{s\in\Delta t_m}$. Because the non-homogeneous Poisson
has independent increments,  it behaves at time $t$ as the sum
of different independent Poisson  processes each with a
constant intensity. The  main consequence of this simple
assumption is that the generation of $y$ at each time step $m$ in Algorithm\myref{alg:spot}, no matter in combination to which methodology illustrated in Subsection\myref{subsect:gou}, is accomplished setting a
different intensity $\lambda_m$ for $m=1,\dots M$.

\subsection{Positive and negative jumps: \bgou process \label{subsect:bgou}}
The three market models presented in Section\myref{sec:markets} all exhibit positive and negative jumps that are modeled as the difference of two \gou\ processes, hence a \bgou\ process. As illustrated in Algorithm\myref{alg:spot}, the generation of the jump component is simply obtained by running one of the algorithms discussed in Subsection\myref{subsect:gou} two times. On the other hand, as shown in Cufaro Petroni and Sabino\mycite{cs20}, one can implement a simulation procedure specific to the process $Y(t)$ with Laplace jumps. In practice, steps four and five of Algorithm\myref{alg:spot} are packed into one because $i=1$.

For instance, the fifth step in Algorithm\myref{alg:ct} has to be replaced by.
\begin{algorithm}
\caption{ }\label{alg:ct:laplace}
\begin{algorithmic}[1]
		\setcounter{ALG@line}{4}
		\State Generate $N$ \iid\ $U_n\sim\erl_1(\lambda_J), n=1, \dots N$, \Comment {Exponential \rv~ with scale $\beta$}
		\State Generate $N$ \iid\ $D_n\sim\erl_1(\lambda_J), n=1, \dots N$, \Comment {Exponential \rv~ with scale $\beta$}
		\State $J_n \gets U_n - D_n, \,n=1, \dots N$
		\end{algorithmic}
\end{algorithm}

In addition, the \chf\ of the process $Y(t)$ at time $t$ is
 \begin{equation}\label{eq:chf:symm:bgou}
\varphi(u, t)=\left(\frac{\beta^2-u^2e^{-2kt}}{\beta^2-u^2}\right)^{\frac{\lambda}{2k}},
\end{equation}
that means that the law of the process at time $t$ coincides with that of the \arem\ $Z_a$ of a symmetrical $\bgam$ with parameters $(\lambda/(2k), \beta)$, taking once again $a=e^{-kt}$.
Algorithm\myref{alg:cs} can then be adapted to the case of a symmetric \bgou\ process as summarized in Algorithm\myref{alg:cs:laplace}.
\begin{algorithm}
\caption{}\label{alg:cs:laplace}
		\begin{algorithmic}[1]		
		\State $\alpha\gets\lambda/2k,\;\; a\gets e^{-k\Delta t_m}$
		\State $b\gets B\sim \pol(\alpha,1-a^2)$ \Comment{Generate a Polya $(\alpha,1-a^2)$ \rv}
		\State $y_i \gets \sim\erl_{b}\left(^{\beta}/_a\right), i=1,2$; \Comment{ Generate two independent  Erlang \rv's with the same rate $\beta/	a$}
		\State $y = y_1 - y_2$
		\end{algorithmic}
\end{algorithm}

Finally, we conclude this subsection noting that the \chf\ in\refeq{eq:chf:symm:bgou} can be rewritten as (see Cufaro Petroni and Sabino \cite{cs20})
\begin{equation}
\varphi(u, t)= e^{\lambda_P t  \left(\varphi_{\tilde{L}}(u)-1\right)}
\end{equation}
where
\begin{equation}
\varphi_{\tilde{L}}(u)=\int_0^1\frac{\lambda_J^2e^{2ktv}}{\lambda_J^2e^{2ktv} + u^2}dv
\label{eq:chf:bgou:qdz}
\end{equation}
\noindent The right-hand side in\refeq{eq:chf:bgou:qdz} is then the \chf\ of compound Poisson whose jumps are independent copies $\tilde{J}_n$ distributed according to a uniform mixture of centered Laplace laws with random parameter $\beta e^{ktU}$ with $U\sim\unif([0,1])$.
This result leads to the adaptation of the methodology of Qu et al.\mycite{QDZ19} to the case of a symmetric \bgou\ detailed in Algorithm\myref{alg:symm:bgou:qdz}.
\begin{algorithm}
\caption{ }\label{alg:symm:bgou:qdz}
\begin{algorithmic}[1]
		\State Generate $N\sim\poiss(\lambda\Delta t_m)$,\Comment {Poisson \rv~with intensity $\lambda\Delta t_m$}
		\State Generate $N$ \iid\ uniform \rv's $\bm{u}=(u_1, \dots, u_N)\,\sim\,\unif([0,1]^N)$.
		\State $\beta_n^{(r)}\gets\beta e^{k\Delta t_m u_n}, n=1,\dots, N, r\in\{u, d\}$.
		\State Generate $N$ \iid\ $U_n\sim\erl_1(\beta_n^{(u)}), i=1,\dots, n$, \Comment {Generate $N$ independent exponential \rv's with random rate $\beta_i^{(d)}}$
		\State Generate $N$ \iid\ $D_n\sim\erl_1(\beta_n^{(d)}), n=1,\dots, N$, \Comment {Generate $N$ independent exponential \rv's with random rate $\beta_n^{(d)}}$
		\State $y\gets \sum_{n=1}^N (U_n - D_n)$		
		\end{algorithmic}
\end{algorithm}

        \section{ Numerical Experiments\label{sect:numExperiments}}
We compare the computational performance of all the algorithms detained in Section\myref{sec:gen:ou} in combination with  Algorithm\myref{alg:spot} for the simulation of the path trajectory of each market model introduced in Section\myref{sec:markets}. We illustrate their differences by pricing  energy
contracts namely, Asian options, swings and storages with Monte Carlo (MC) methods. The implementation of the pricing of such contracts
with MC methods needs to be unbiased and fast especially if it is meant for real-time calculations.

In our numerical experiments, we decided to assign different mean-reversion rates
to the jump and to the diffusive components to
better capture  the spikes.
For example, with respect to the parameter settings used in
Deng\cite{Deng00stochasticmodels} and Kjaer\cite{Kjaer2008}, the
mean-reversion rates of  our jump components are
larger than those of their diffusion counterparts. The parameter
combination in Kjaer\cite{Kjaer2008} assumes  indeed that
the process $H(t)$ has  just one -- and small --
mean-reversion rate with a high $\lambda$,  so that
$\lambda/k\simeq 7$ and one could implement the simplified
version of Algorithm\myref{alg:cs} based on the binomial mixture of Erlang laws with $\alpha$ being an
integer number as explained in Cufaro Petroni and Sabino\mycite{cs20}.

All the simulation experiments in the present paper
have been conducted using \emph{MATLAB R2019a} with a $64$-bit
Intel Core i5-6300U CPU, 8GB.
As an additional validation, the
comparisons of the simulation computational times have
also been performed with \emph{R} and \emph{Python}  leading to the same conclusions.

        \subsection{ Numerical Experiments: Asian Options \label{subsect:numExperiments:asian}}		
The first numerical experiment that we have conducted, refers to the pricing of an Asian option with European exercise style using MC under the assumption that the jump process $Y(t)$ of the market model\refeq{eq:spot} is given by\refeq{eq:spot:jump:kou} (case 1). Therefore, it results
\begin{equation*}
	h(t) = -\frac{\sigma_D^2}{4k_D}\left(1-e^{-2k_Dt}\right) - \frac{p\lambda}{k}\log\left(\frac{\beta_1 - e^{-k t}}{\beta_1 - 1}\right)
	-\frac{(1-p)\lambda}{k}\log\left(\frac{\beta_2 - e^{-kt}}{\beta_2 - 1}\right).
\end{equation*}
Recalling that the payoff of such an option at maturity $T$ is
\begin{equation*}
	A(T) = \left(\frac{\sum_{i=1}^{M}S(t_i)}{M} - K\right)^+,
\end{equation*}
we consider an at-the-money Asian option $K=S_0=22$ having one year maturity ($T=1$) and with realistic market parameters shown in Table~\ref{tab:spot:kou} with a flat forward
curve.

Although the calibration is not the focus of this paper, the market parameters can be considered realistic (they are comparable to those in Kjaer\mycite{Kjaer2008} or Deng\mycite{Deng00stochasticmodels}). In addition, we remark
that in Cufaro Petroni and Sabino\mycite{cs20}, we have found the transition density of the \gou\ and \bgou\ processes in close form. Therefore, this
gives the possibility (at least in terms of convolution) to write
 down the overall transition density, and hence the likelihood
function. As an alternative, one could also apply one of the estimation procedures illustrated in Barndorff-Nielsen and Shephard
\cite{BNSh01} with the advantage that the eventual estimated parameters would
not be affected by the  approximations implicit in any
discretization scheme (besides truncating the infinite
series).
\begin{table}
				\caption{Parameters for Spot (day-ahead) dynamics (Case 1).} \label{tab:spot:kou}
				\centering
						\begin{tabular}{|c|c|c||c|c|c|c|c|c|}
							\hline
								 $S_0$ & $k_D$ & $\sigma_D$ & $k$ &  $p$ & $\lambda_1$ & $\lambda_2$ & $\beta_1$ & $\beta_2$ \\
							\hline
								 $22$ & $67$ &  $0.25$ & $50$ & $50$ & $0.6$ & $20$ & $10$ & $20$ \\
							\hline
						\end{tabular}
\end{table}
\input{./Tables/Asian}

\begin{figure}
\caption{Asian options.}\label{fig:comp:times:asian_gbm_gauss_laplace}
		\begin{subfigure}[c]{.5\textwidth}{
				\includegraphics[width=70mm]{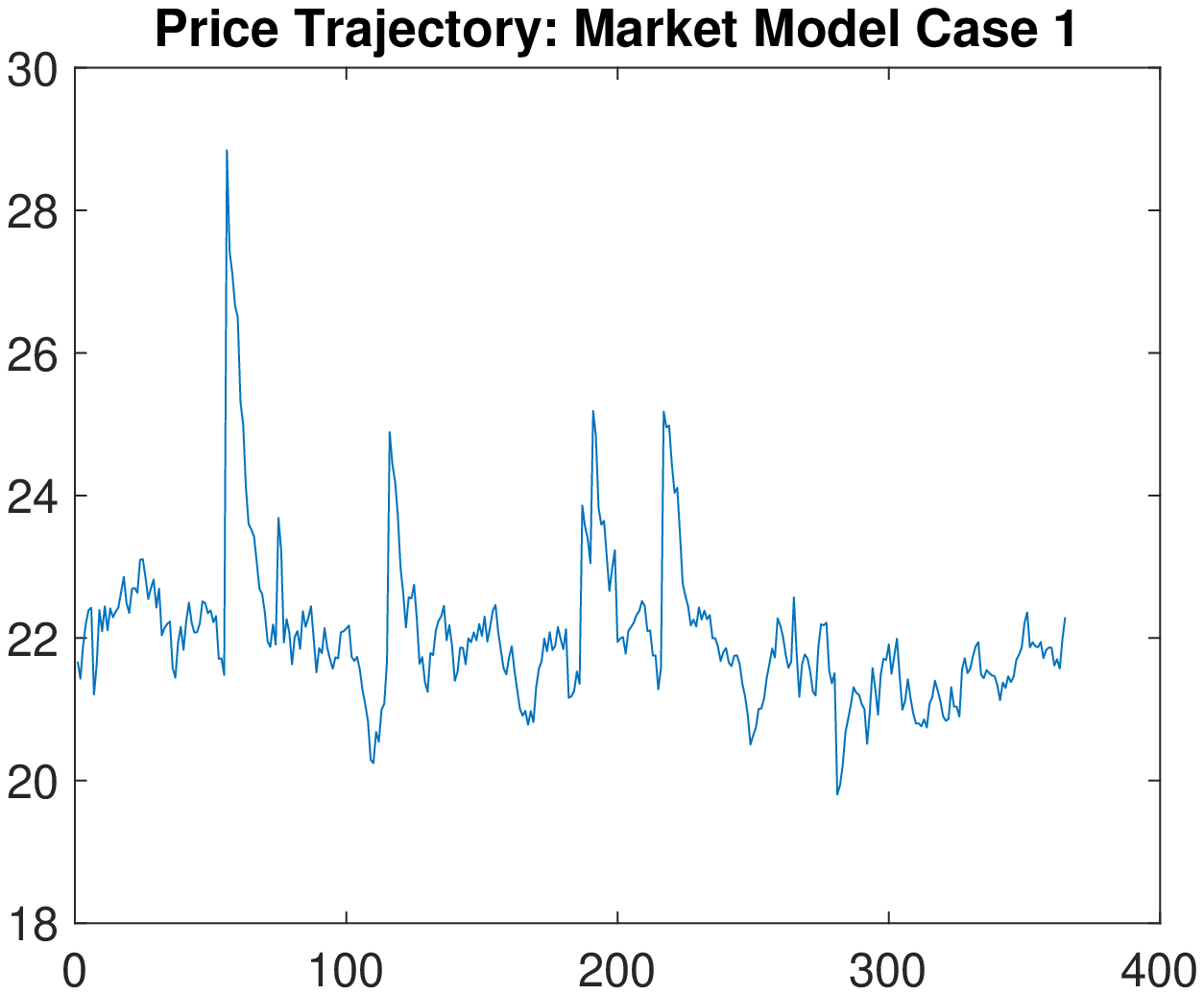}
				}
				\caption{Sample Trajectory.}\label{fig:trajectory:asian}
		\end{subfigure}
		\begin{subfigure}[c]{.5\textwidth}{
				\includegraphics[width=70mm]{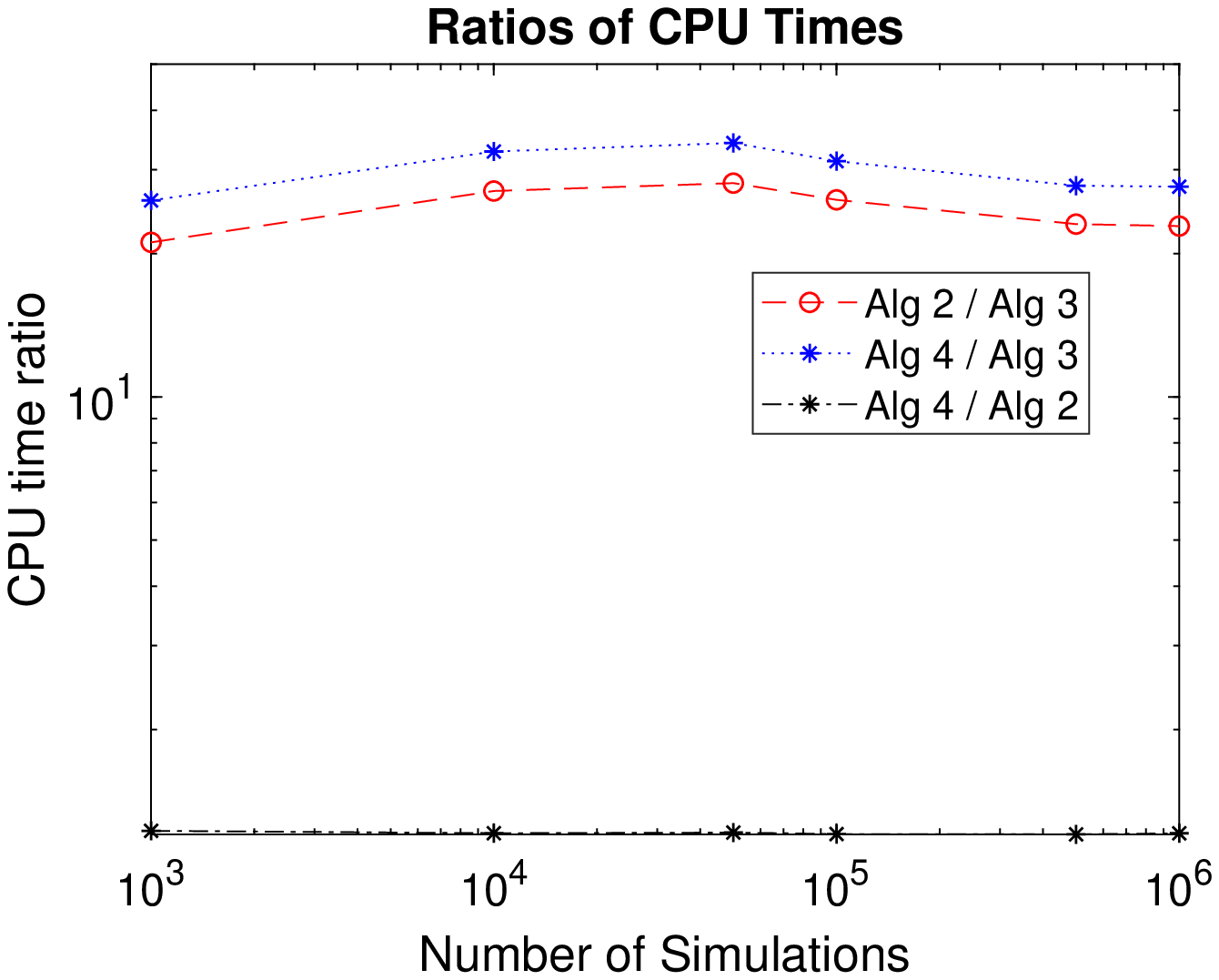}
				}
				\caption{Ratio of CPU times.}\label{fig:ratio:comp:times:asian}
		\end{subfigure}
\end{figure}%
Table~\ref{tab:Asian:MC} shows the estimated prices, the RMSE's (divided by the squared root of the number of simulations) and the overall CPU times in seconds using the different methodologies for the simulation of the process $Y(t)$. As expected, in terms of convergence, all the approaches are equally performing while instead, the CPU times are radically different. Algorithm\myref{alg:ct} and Algorithm\myref{alg:qdz} have similar computational effort therefore their CPU times are comparable as  observed in Figure~\ref{fig:ratio:comp:times:asian}. On the other hand, our methodology provides a remarkable computational advantage: what requires minutes for the Algorithms\myref{alg:ct} and\myref{alg:qdz} only requires seconds for the Algorithm\myref{alg:cs}. For example with $N=5\times 10^5$ simulations, with our computer, the pricing of the Asian option above,  is accomplished in almost two minutes whereas, it takes almost one hour with the other alternatives. Figure~\ref{fig:ratio:comp:times:asian} clearly shows that, in the worst case, our simulation procedure is at least thirty times faster than any other alternative being then suitable for real-time applications.

				
        \subsection{ Numerical Experiments: Gas Storages \label{subsect:numExperiments:gasstorage}}
Denote by $C(t)$ the volume of a (virtual) gas storage at time $t$
with $C_{min}\le C(t)\le C_{max}$. The holder of such an energy
asset is faced with a timing problem that consists  in
deciding when to inject, to withdraw or to do-nothing.

 Denoting $J(t,x,c)$ the value of a gas storage at time $t$
given $S(t)=x$, $C(t)=c$, one can write:
            \begin{equation}\label{eq:LSMC}
                J(t,x,c) = \sup_{u\in\mathcal{U}}\mathbb{E}\left[\int_t^T \phi_u\left(S(s) \right)ds + q\left(S(T),C(T) \right)\,\right| S(t)=x, C(t)=c\bigg],
            \end{equation}
 where $\mathcal{U}$ denotes the set of  the admissible
strategies, $u(t)\in\{-1,0,1\}$ is the regime at time $t$ such that
            \begin{equation}
            \left\{
                \begin{array}{lcll}
                     \phi_{-1}(S(t)) &=& -S(t)-K_{in} a_{in}, & \text{injection} \\
                    \phi_{0}(S(t)) &=& -K_N, & \text{do nothing} \\
                    \phi_{1}(S(t)) &=& S(t)-K_{out} a_{w} &\text{withdrawal}
                \end{array}
            \right.,
            \end{equation}
$a_{in}$ and $a_{w}$ are the injection and withdrawal rates,
$K_{in}$, $K_{out}$ and $K_N$,  respectively, represent the
costs of injection, do-nothing and withdrawal,  and $q$ takes
into account the possibility of final penalties.
Based on the Bellman recurrence equation (see Bertsekas\mycite{Bertsekas05}), one can perform the following backward recursion for $i=1,\dots,d$:
                \begin{equation}
                    J(t_i,x,c) = \sup_{k\in\{-1,0,1\}} \left\{\phi_k S(t_i) + \mathbb{E}\left[ J\left(t_{i+1},S(t_{i+1}),\tilde{c}_k\right)| S(t_i)=x, C(t_i)=c\right]   \right\} , i=1,\dots,d,
                \end{equation}
                where
                \begin{equation}
                \left\{
                \begin{array}{lll}
                    \tilde{c}_{-1} &=& \min(c+a_{in}, C_{max})\\
                    \tilde{c}_{0} &=& c\\
                    \tilde{c}_{1} &=& \min(c-a_w, C_{min}).\\
                \end{array}
                \right.
                \end{equation}
A standard approach to price gas storages is a modified
version of the Least-Squares Monte Carlo (LSMC), introduced in
Longstaff-Schwartz\mycite{LSW01}, detailed in Boogert and de
Jong\mycite{BDJ08}. With this approach, the backward recursion is obtained by defining a
finite volume grid of G steps for the admissible capacities $c$ of the plant and then apply
the LSMC methodology to the continuation value per volume step. In alternative, one may solve the recursion by adapting the method proposed by  Ben-Ameur et al.\mycite{BBKL2007} or might use the quantization method as explained in Bardou et al.\mycite{BBP07}.
Although the LSMC might not be the fastest solution, risk management units of energy companies are often interested in quantiles of the price distribution that can be obtained as a side product using the LSMC method.

We focus then on the LSMC methodology and perform  a few numerical experiments selecting the
three-factors spot model with the jump component covered by the second case in Section\myref{sec:markets}
because we want to capture asymmetric jumps (we set $H(0)=0)$: in this
case, because of\refeq{eq:rn:spot} and\refeq{eq:chf:gou} for $\beta_1,\beta_2>1$
it results
            \begin{equation*}
              h(t) = -\frac{\sigma_D^2}{4k_D}\left(1-e^{-2k_Dt}\right) - \frac{\lambda_1}{k_1}\log\left(\frac{\beta_1 - e^{-k_1 t}}{\beta_1 - 1}\right)
              -\frac{\lambda_2}{k_2}\log\left(\frac{\beta_2 - e^{-k_2t}}{\beta_2 - 1}\right).
            \end{equation*}
            \begin{table}
                    \caption{Parameters for Spot (day-ahead) dynamics (Case 2)} \label{tab:spot3Factor}
                    \centering
                        \begin{tabular}{|c|c|c||c|c|c|c|c|c|}
                          \hline
                             $S_0$ & $k_D$ & $\sigma_D$ & $k_1$ &  $k_2$ & $\lambda_1$ & $\lambda_2$ & $\beta_1$ & $\beta_2$ \\
                          \hline
                             $22$ & $67$ &  $0.25$ & $50$ & $40$ & $20$ & $20$ & $10$ & $20$ \\
                          \hline
                        \end{tabular}
            \end{table}

This model can also be extended to cover
correlated Poisson processes. For instance, in Cufaro Petroni and
Sabino\mycite{cs17} and\mycite{CufaroSabino:QF18}  we once more used
the concept of \sd\ to  produce correlated Poisson processes
with a time-delay mechanism among jumps, and we discussed an
application to the pricing of spread options.  Nevertheless, in this study we consider independent Poisson processes only.

Going back to the initial problem, we assume that the units of
$C(0), C(T)$ and $C_{max}$ are in MWh, those of the injection and
withdrawal rates are in  MWh/day, whereas  $S_0$ can be taken in
\text{\euro}/MWh; in addition we suppose a flat forward
curve. The remaining model parameters are shown in
Table\myref{tab:spot3Factor} and can be considered realistic.

\begin{figure}\label{fig:gas:storages}
                \caption{Gas Storages.}
                    \begin{subfigure}[c]{.5\textwidth}{
                      \includegraphics[width=70mm]{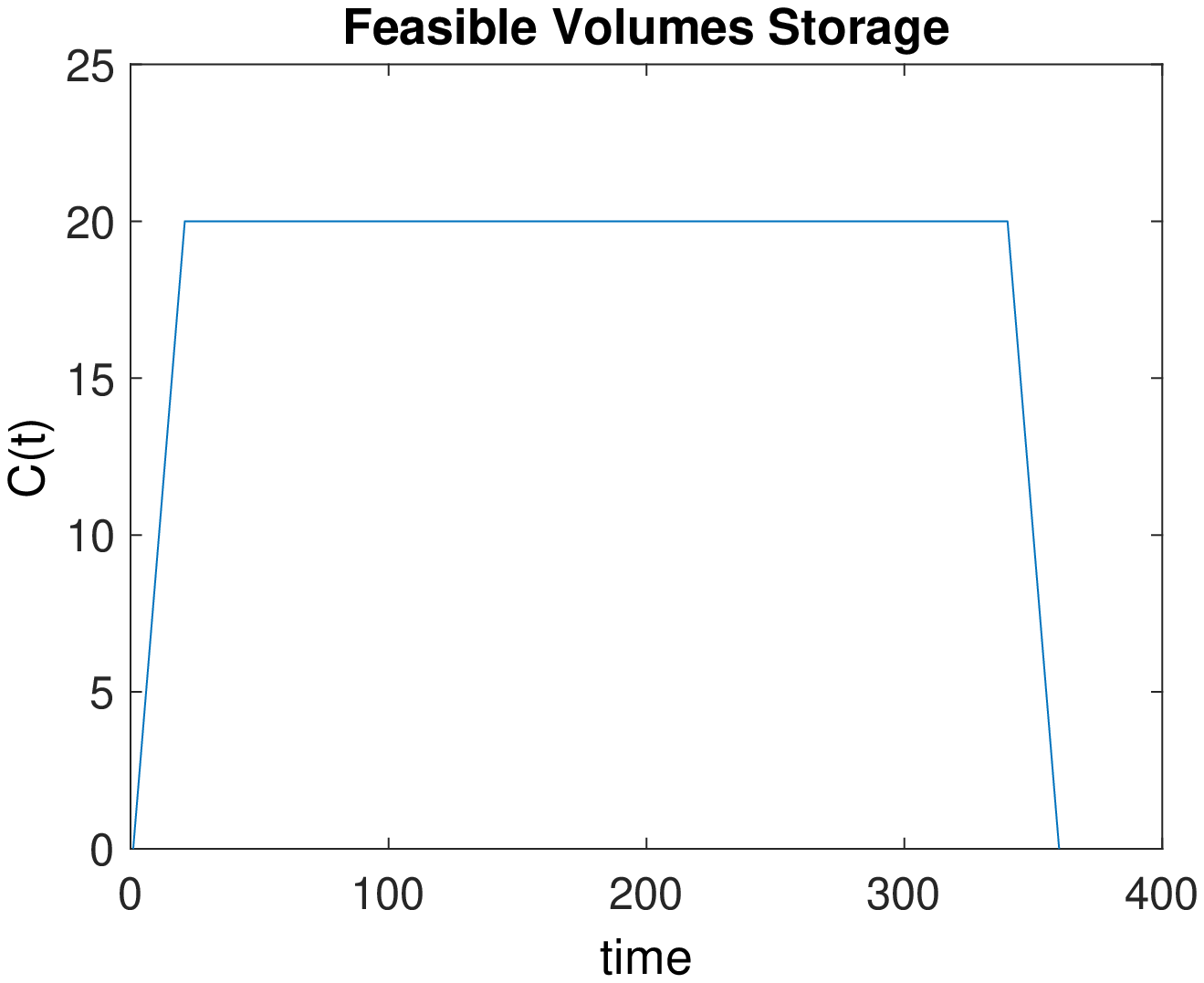}
                      \caption{Feasible Volumes of the fast churn storage}\label{fig:fast_churn_storage}
                      }
                    \end{subfigure}
                    \begin{subfigure}[c]{.5\textwidth}{
                      \includegraphics[width=70mm]{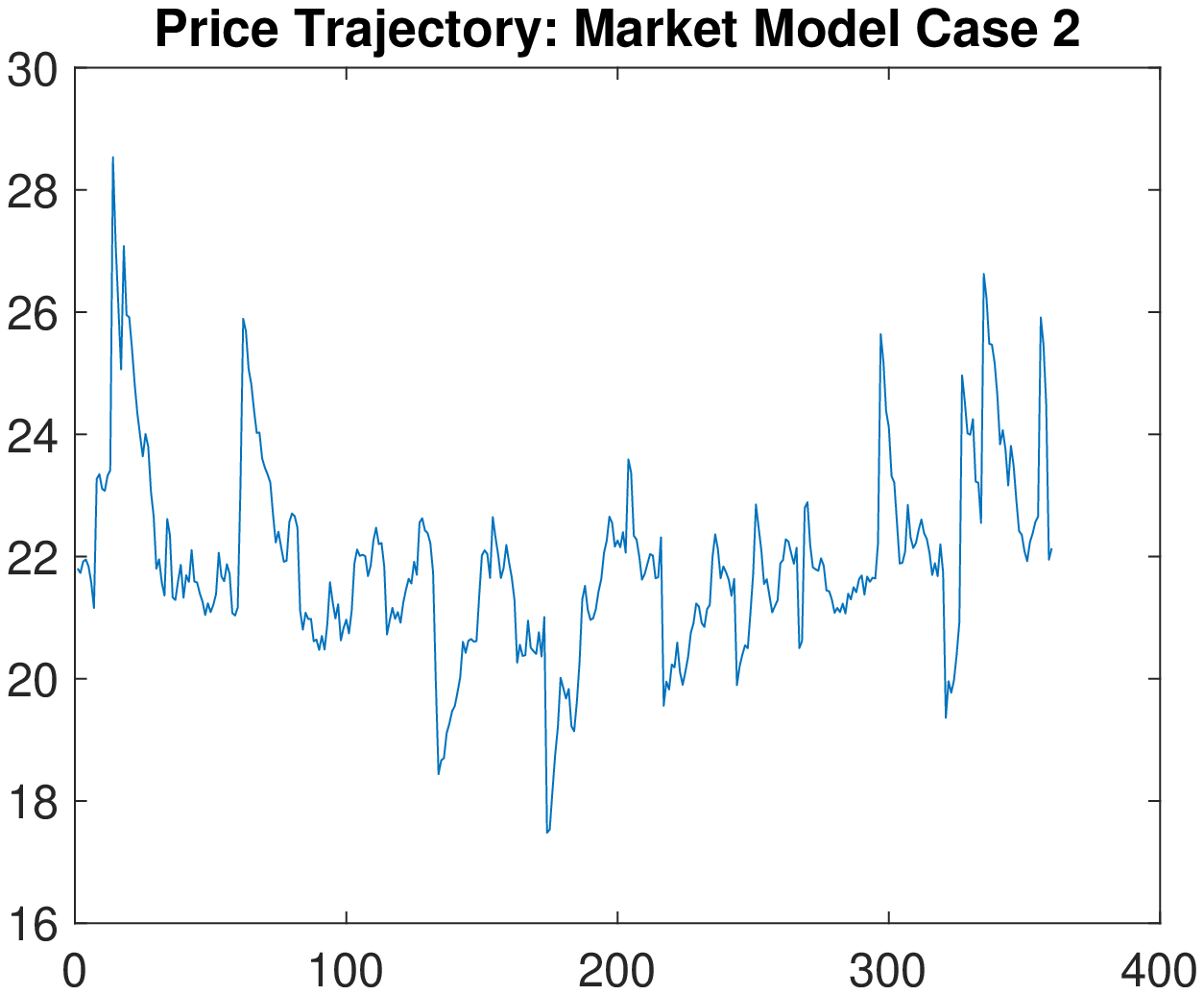}
                      \caption{Sample Trajectory}\label{fig:trajectory:gas:storage}
                      }
                    \end{subfigure}
                \end{figure}

We consider  finally a one-year fast-churn storage with the
parameters shown in Table\myref{tab:storage:spec} such that $20$
days are required to fill or empty the storage as shown in
Figure\myref{fig:fast_churn_storage}.
            \input{./Tables/TableGasStorageSpec}
						\input{./Tables/Storage}

            \begin{figure}\label{fig:gas:storage:results}
            \caption{Gas Storage Results.}
                \begin{subfigure}[c]{.5\textwidth}{
                  \includegraphics[width=70mm]{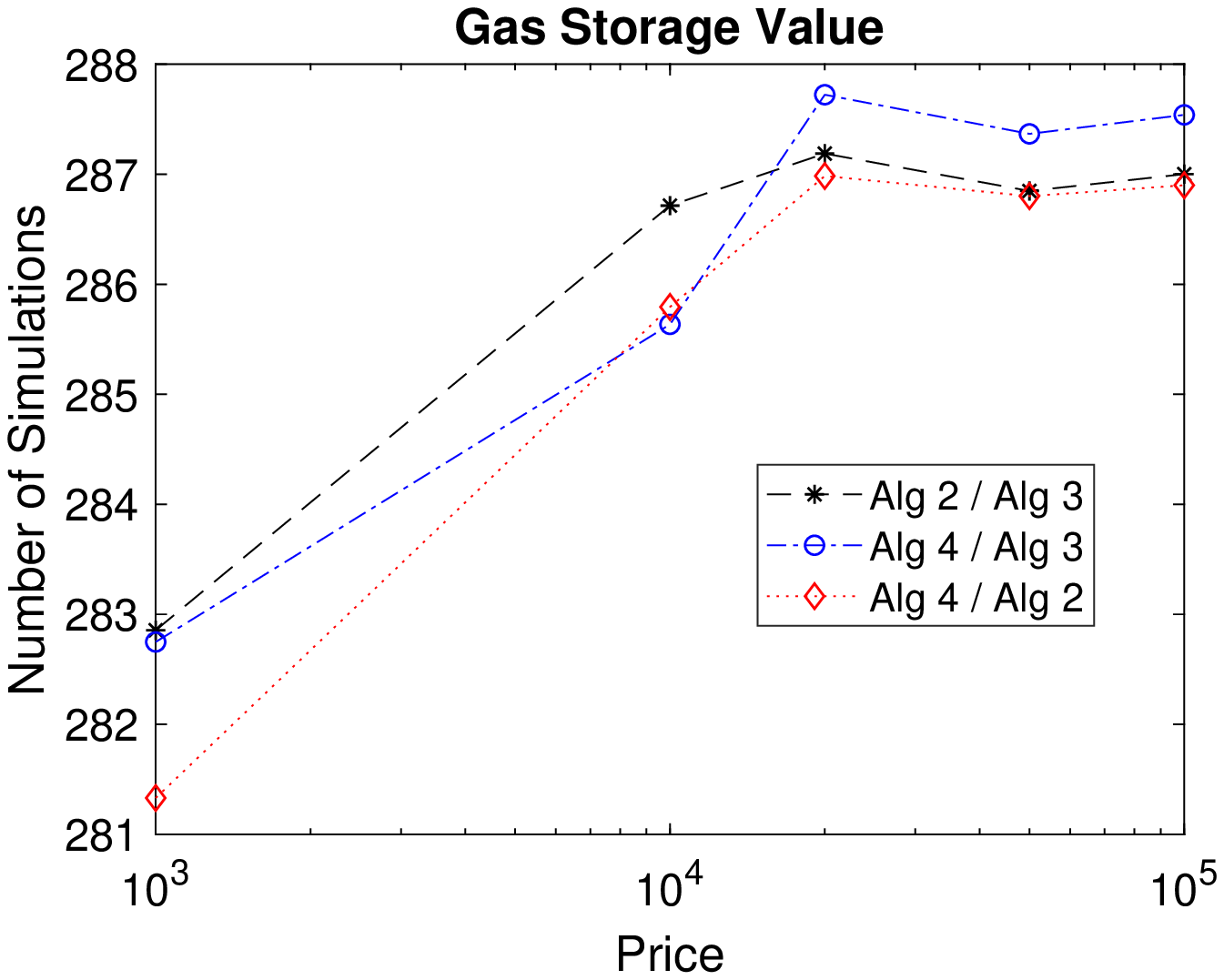}
                  \caption{Values Gas Storage.}\label{fig:value:storage}
                  }
                \end{subfigure}
                \begin{subfigure}[c]{.5\textwidth}{
                  \includegraphics[width=70mm]{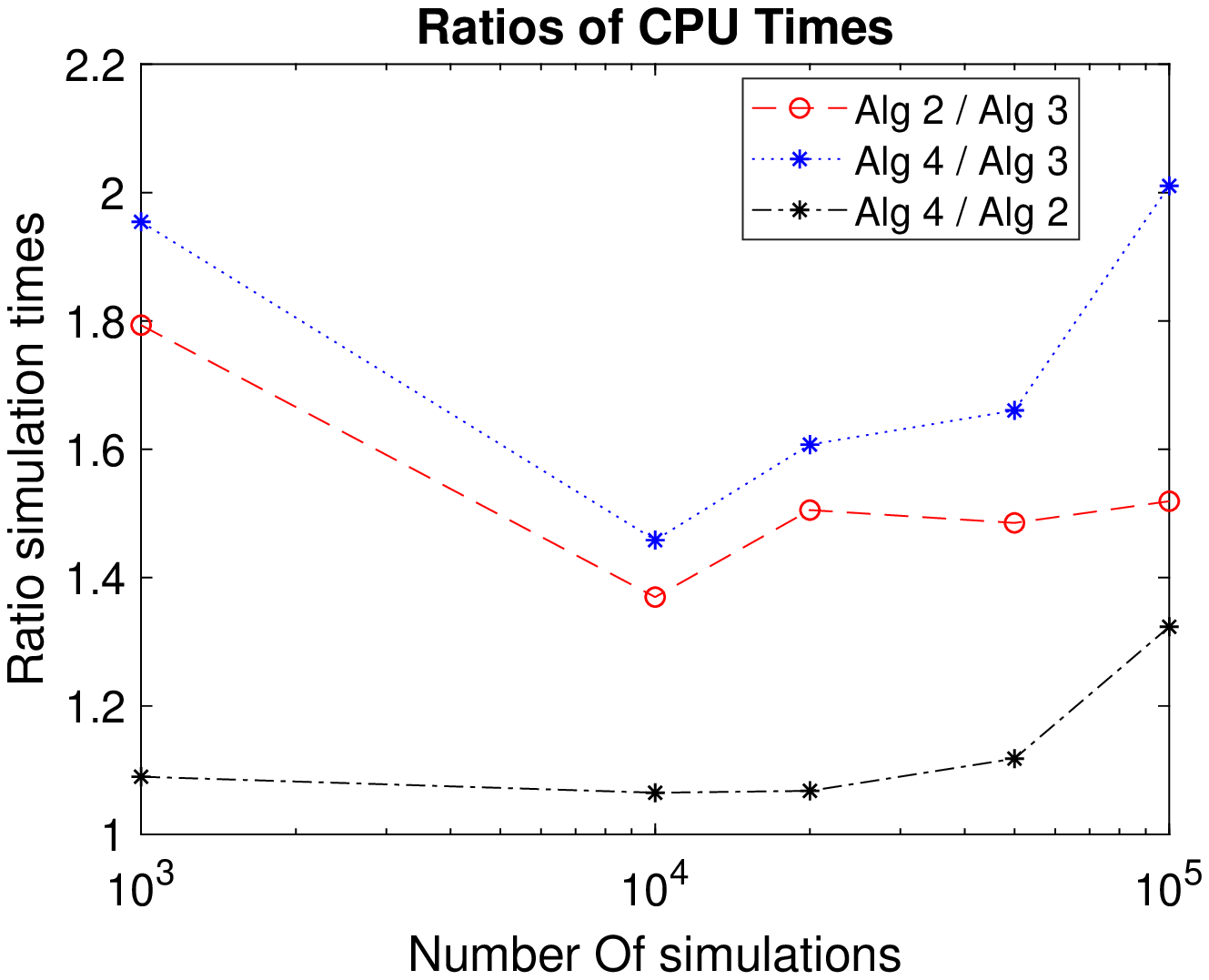}
                  \caption{Ratio of CPU Times.}\label{fig:times:storages}
                  }
                \end{subfigure}
            \end{figure}%
In line with that observed for the pricing of Asian options, Table\myref{tab:Storage:MC} and Figure\myref{fig:value:storage} show that the three
types of implementation  apparently return comparable gas
storage values. On the other hand, the ratio of overall CPU times in Figure\myref{fig:times:storages} is not as extreme compared to the Asian option case. Algorithm\myref{alg:cs} is ``only''  $40\%$ faster, in the worst case, compared to the other two solutions. The reason of this apparent different conclusion compared to the previous section is that the main component to the overall computational cost derives from the stochastic optimization. To this end, Table\myref{tab:Storage:MC} also displays the CPU times required for the path simulation only (denoted PATH) where one can observe the Algorithm\myref{alg:cs} is once again tens of times faster. Using Algorithms\myref{alg:ct} and\myref{alg:qdz} the path simulation time is a relevant portion of the overall time, whereas using our approach it is, as if the overall cost coincides with that required by the purely LSMC stochastic optimization. This fact provides a computational advantage when one needs to calculate the sensitivities of the storage because a high number of simulations is required.

We finally remark that Algorithm\myref{alg:spot}  relies on the sequential simulation of
the price trajectory forward in time. In combination with LSMC
methods, this is not the optimal approach  because the entire set of
trajectories and simulations are stored in memory with a risk of
memory allocation issues. For instance, Pellegrino and
Sabino\mycite{PellegrinoSabino15} and Sabino\mycite{Sabino20}
have shown that the backward simulation is preferable with LSMC.
Unfortunately,  although we know the law of the standard
Gaussian-\ou\ bridge, we do not know the law of the
\gou\ bridge which will be one of the  topics of
 our future studies.
                    \subsection{ Numerical Experiments: Swings \label{subsect:numExperiments:swings}}

\begin{figure}\label{fig:swing:Results}
\caption{Swings.}
				\begin{subfigure}[c]{.5\textwidth}{
						\includegraphics[width=70mm]{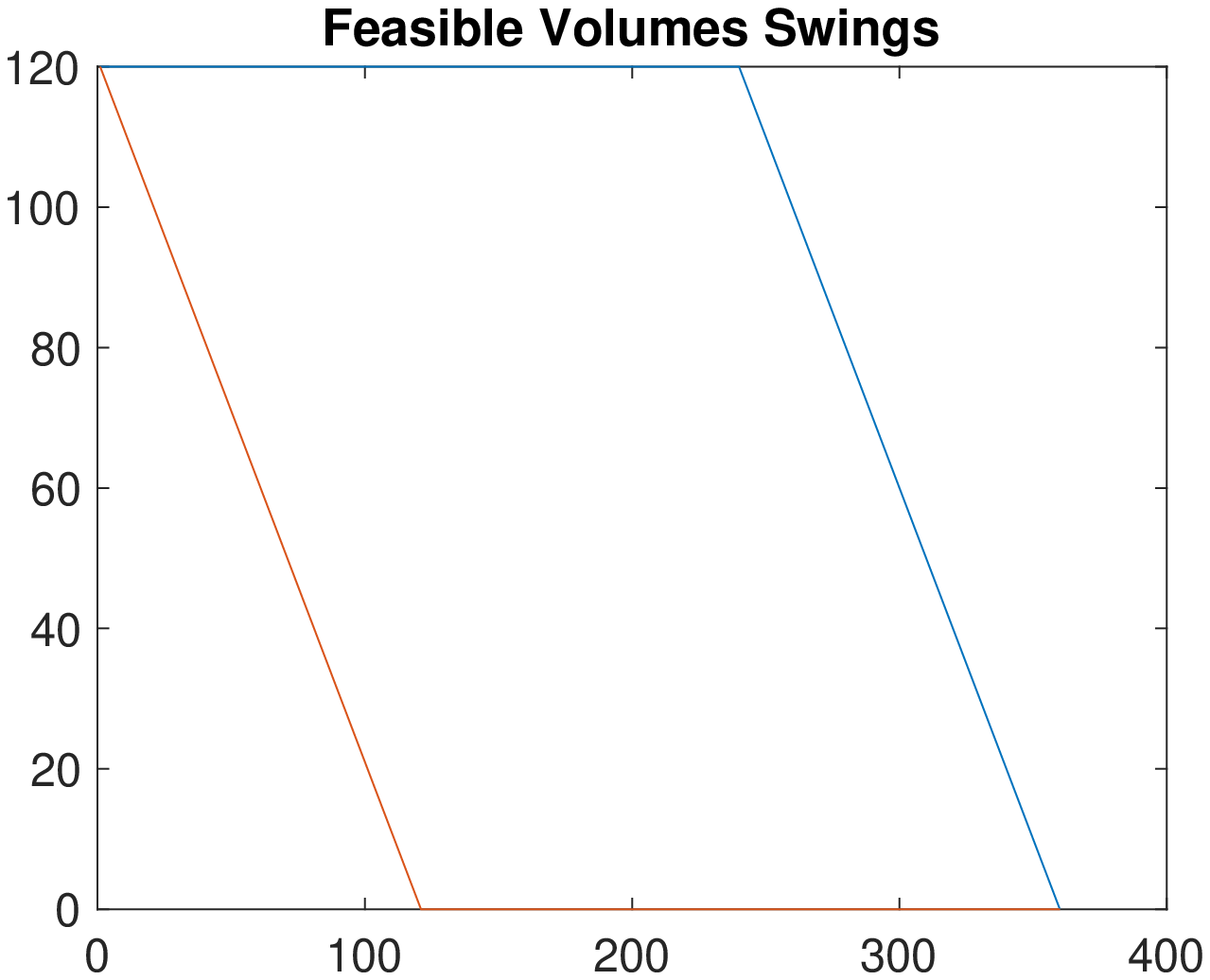}
						\caption{Feasible Volumes of a $120$-$120$ Swing.}\label{fig:swing:volumes}
						}
				\end{subfigure}
				\begin{subfigure}[c]{.5\textwidth}{
						\includegraphics[width=70mm]{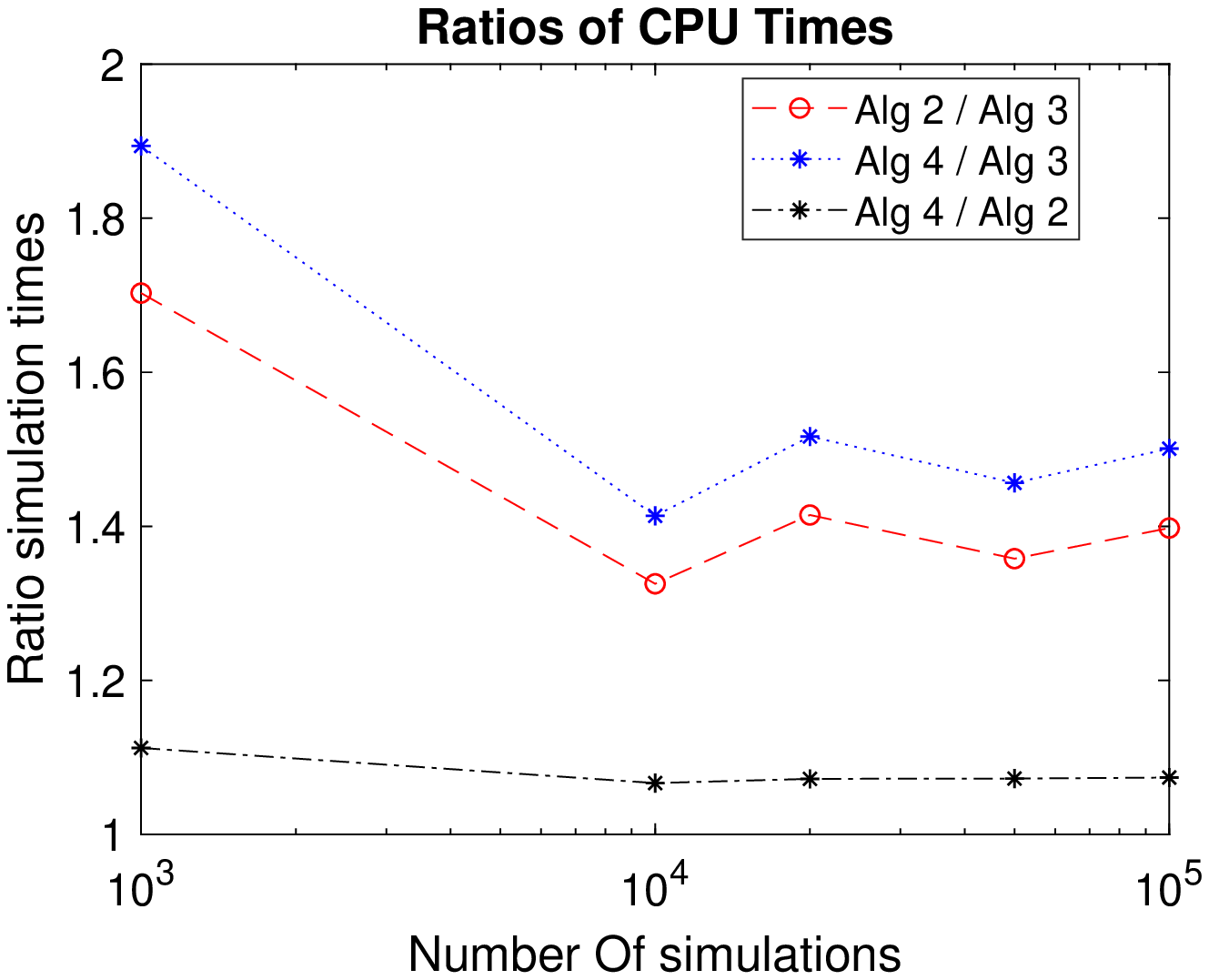}
						\caption{Swing Values and Ratio of CPU times.}\label{fig:times:swings}
						}
				\end{subfigure}
\end{figure}

A swing option is a type of contract used by investors in energy
markets that lets the option holder buy a predetermined quantity of
energy at a predetermined price (strike), while retaining a certain
degree of flexibility in  both the amount purchased and the
price paid. Such a contract can also be seen as a simplified gas
storage  where $a_{in}=0$, $K_N=0$ and $K_w$ is the strike of
the contract. We consider a $120$-$120$ swing  with  the
specifications  of Table\myref{tab:swing:spec} and
Figure\myref{fig:swing:volumes}: it can be seen as plugging
$C(0)=120$, $C(T)=0$, $a_{in}=0$, $a_w=1$, $C_{max}=120$
into\refeq{eq:LSMC} with an injection cost equal to the strike.

In this last example, we
now choose the third market model in Section\myref{sec:markets} that consists in a two-factors model
with  one Gaussian \ou\ diffusion and  one symmetric \bgou\ process - a compound Poisson with Laplace jumps - where once more we set $H(0)=0$. We also
consider a step-wise daily approximation of the following
time-dependent intensity
                                                \begin{equation}\label{eq:td:intensity}
                                                    \lambda(t) = \frac{2\theta}{1 + |sin\left(\pi\omega(t-\tau)\right)|}
                                                \end{equation}
so that for $m=1,\dots, M$ and $\beta>1$  we have
		\begin{equation*}
				h(t_m) = \frac{\sigma_D^2}{4k_D}\left(1-e^{-2k_Dt_m}\right) -\frac{\lambda_m}{2k}\log\left(\frac{\beta^2 - e^{-2kt_m}}{\beta^2 -1}\right)
		\end{equation*}
\noindent with the parameters  of Table\myref{tab:spot2Factor}
once again with a flat forward curve.  The value of $\theta$
is such that the average number of jumps per year is about
$40$ as in the storage example.
\begin{table}
	\caption{Parameters for Spot (day-ahead) dynamics (Case 3)}\label{tab:spot2Factor}
		\centering
						\begin{tabular}{|c|c|c||c|c|c|c|c|}
								\hline
										 $S_0$ & $k_D$ & $\sigma_D$ & $k_N$ & $\theta$ & $\omega$ & $\tau$ & $\beta$ \\
								\hline
										 $22$ & $67$ &  $0.25$ & $50$ & $32$ & $2$ & $0.25$ & $20$ \\
								\hline
						\end{tabular}
\end{table}
\input{./Tables/TableGasSwingSpec}
\input{./Tables/Swing}

\begin{figure}\label{fig:swing:market:model}
\caption{Market Model.}
				\begin{subfigure}[c]{.5\textwidth}{
						\includegraphics[width=70mm]{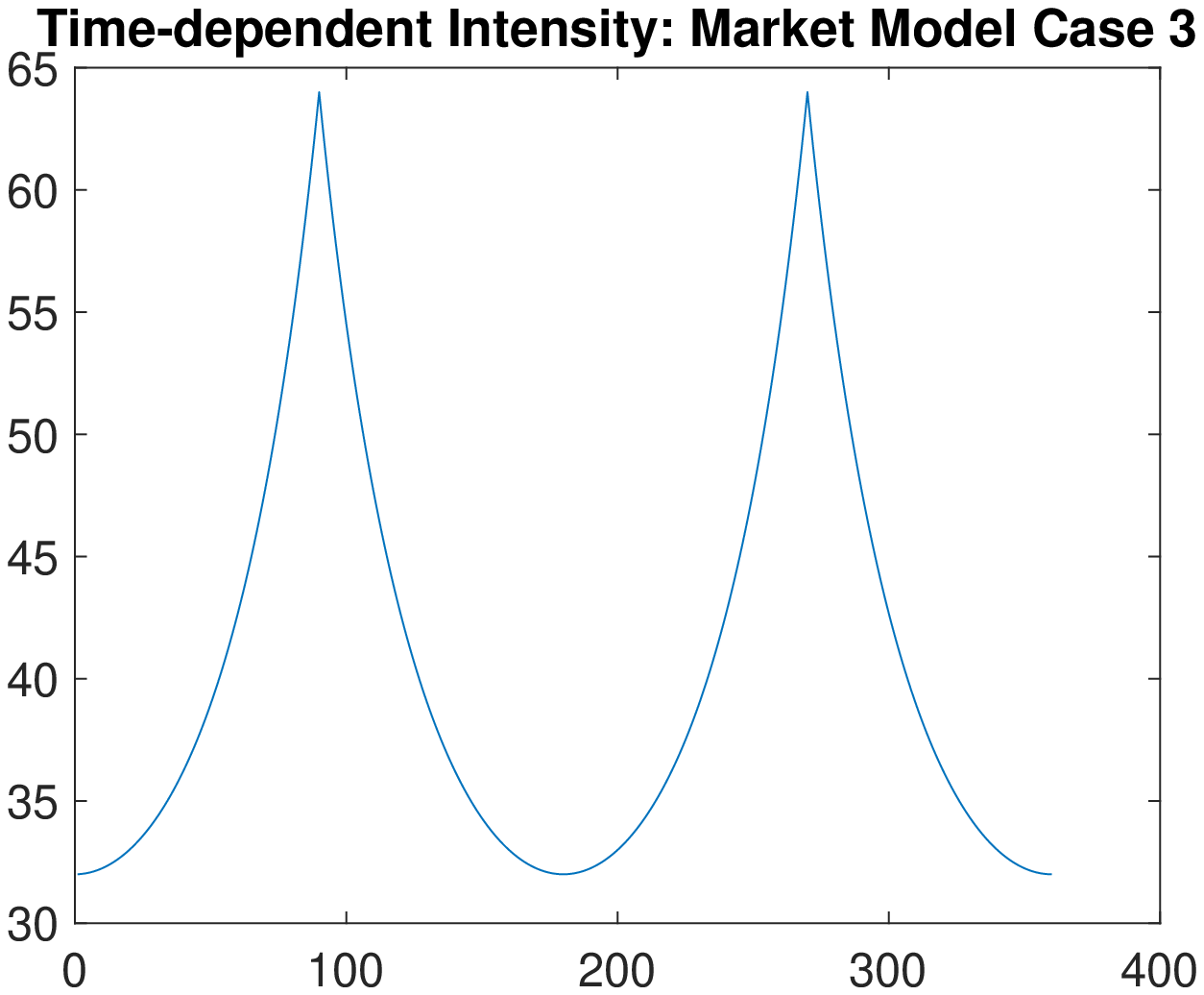}
						\caption{Time-dependent Intensity}\label{fig:time_dependent_intensity}
						}
				\end{subfigure}
				\begin{subfigure}[c]{.5\textwidth}{
						\includegraphics[width=70mm]{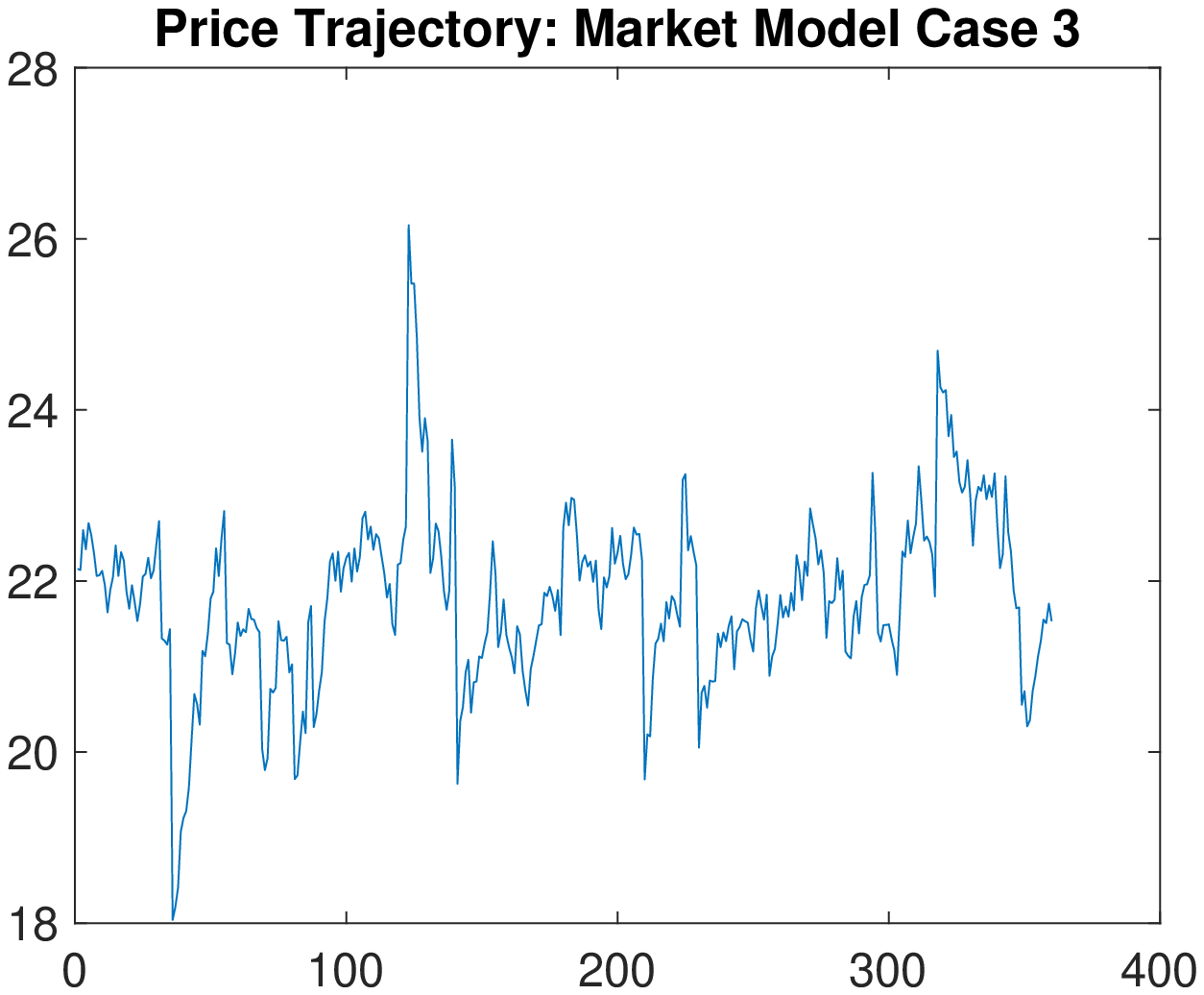}
						\caption{Sample Trajectory.}\label{fig:trajectory_swing}
						}
				\end{subfigure}
\end{figure}%
Due to the fact that jump component has now symmetric Laplace jumps, the process $Y(t)$ can be seen as a symmetric \bgou\ process therefore, instead of executing Algorithms\myref{alg:ct},\myref{alg:cs} and\myref{alg:qdz} two times, we can rely on Algorithms\myref{alg:ct:laplace},\myref{alg:cs:laplace} and\myref{alg:symm:bgou:qdz}.
The conclusions that we can derive from the
numerical experiments are very much in line with what is observed in
the case of gas storages. As expected, the MC based estimated values of the swing option obtained with the three
types of implementation are similar.

As shown in Table\myref{tab:Swing:MC} and Figure~\ref{fig:times:swings}, once more the
CPU times  with Algorithm\myref{alg:cs:laplace} are far
lower resulting in a competitive advantage of  about $40\%$ (in the worst case) on
the overall computational cost (LSMC in Table\myref{tab:Swing:MC}). This factor becomes even higher if
one focuses on the time required to simulate the price paths (PATH in Table\myref{tab:Swing:MC}). The
contribution of the stochastic optimization step to the overall
cost is again of about $75\%$ using Algorithm
\ref{alg:ct:laplace} or Algorithm\myref{alg:symm:bgou:qdz}, while instead, with Algorithm\myref{alg:cs:laplace}, the path generation step becomes almost
negligible compared to the total CPU time. We can therefore conclude that Algorithm\myref{alg:cs:laplace} is the preferable solution for the simulation of the jump component in the market model\refeq{eq:spot}.

        \section{ Conclusions and future inquiries\label{sec:conclusions}}
In this paper we have considered the problem of pricing complex energy derivatives with Monte Carlo simulations using mean-reverting jump-diffusion market models.
The jump component that we have chosen is a compound Poisson process with exponentially or bilateral exponentially distributed jumps known in the literature as \gou\ or \bgou\ processes.
Although, this is a simple and standard approach, the simulation of the price trajectories may soon become very computational expensive, especially for the pricing of complex derivative contracts.
Indeed, the generation of the path of the jump process has a relevant impact on the overall computational cost.

Based on our results in Cufaro Petroni and Sabino\mycite{cs20},  the main contribution of this paper is the design of exact and very fast simulation algorithms for the simulation of the spot prices that potentially could be used for real-time pricing.

We illustrated the applications of our findings in the
context of the pricing of Asian options with standard Monte Carlo and gas storages and swings adopting the
Least-Squares Monte Carlo method introduced in Boogert and de
Jong\cite{BDJ08}.
The overall computational effort depends on the cost of
simulating the price trajectories and the stochastic optimization
 (this last step is not influenced by the particular simulation
algorithm).

We have conducted extensive simulation experiments and compared the performance of our proposal to the traditional approach of Cont and Tankov\mycite{ContTankov2004} and a recent methodology described by Qu et al.\mycite{QDZ19}. Our numerical experiments have shown that our
solution outperforms any other alternative, because it cuts the
simulation time down by a factor larger than forty in the case of Asian options and to a factor of forty percent for the gas storages and swings. In contrast to the other approaches, the numerical tests suggest that our simulation methodology is suitable for real-time pricing.

From a mathematical point of view,
it would be interesting to study if --  and under which
conditions --  our results could be generalized to other Ornstein-Uhlenbeck processes used in financial applications and in
energy markets (see for instance Cummins et al.\cite{CKM17}).

In a primarily economic and financial perspective, the future
studies could cover the extension to  a multidimensional
setting with correlated Poisson processes as those introduced
for instance in Lindskog and McNeil\mycite{LindskogMcNeil} or in
Cufaro Petroni and Sabino\mycite{cs17}. A last topic deserving
further investigation is a possible enhancement of the
computational speed relying on backward simulations
generalizing the results of Pellegrino and
Sabino\mycite{PellegrinoSabino15} and Sabino\mycite{Sabino20} to the
case of \gou\ or \bgou\ processes.

	\bibliographystyle{plain}
   \bibliography{Cufaro_Sabino_Fast_Pricing}
\end{document}

%% file: Tables/Asian.tex
\begin{table}[ht!]
    \centering\footnotesize
		\resizebox{\textwidth}{!}{
        \begin{tabular}{*{10}{|c|ccc|ccc|ccc}}
					\hline	
										\multicolumn{1}{|c}{} & \multicolumn{3}{|c}{Algorithm \ref{alg:cs}} & \multicolumn{3}{|c}{Algorithm \ref{alg:ct}} & \multicolumn{3}{|c|}{Algorithm \ref{alg:qdz}}\\
					\hline
$N$ & $\text{Price} $ & $\text{RMSE}$ & $\text{CPU}$ & $\text{Price} $ & $\text{RMSE}$ & $\text{CPU}$ & $\text{Price} $ & $\text{RMSE}$ & $\text{CPU}$\\
					\hline
$10^3$ & $1.0299$ & $0.0507$ & $0.29$ & $1.0712$ & $0.0525$ & $6.12$ & $1.0404$ & $0.0503$ & $7.49$\\
$10^4$ & $0.9987$ & $0.0160$ & $2.12$ & $0.9887$ & $0.0159$ & $57.45$ & $1.0131$ & $0.0160$ & $69.56$\\
$5\times 10^4$ & $1.0054$ & $0.0072$ & $10.20$ & $1.0198$ & $0.0072$ & $286.85$ & $1.0219$ & $0.0073$ & $348.39$\\
$10^5$ & $1.0155$ & $0.0051$ & $22.14$ & $1.0126$ & $0.0051$ & $574.44$ & $1.0140$ & $0.0051$ & $692.70$\\
$5\times 10^5$ & $1.0148$ & $0.0023$ & $124.90$ & $1.0191$ & $0.0023$ & $2881$ & $1.0106$ & $0.0023$ & $3469$\\
$10^6$ & $1.0172$ & $0.0016$ & $251.45$ & $1.0153$ & $0.0016$ & $5745$ & $1.0178$ & $0.0016$ & $6953$\\
					\hline
        \end{tabular}				
		}
    \scriptsize
    \caption{\footnotesize{Asian Options: CPU times in seconds and comparison among the prices and RMSE's obtained with the different algorithms}}\label{tab:Asian:MC}
\end{table}


%% file: Tables/TableGasStorageSpec.tex
\begin{table}
    \centering
    \begin{tabular}{|c|c|c|c|c|}
            \hline
             $C(0) $   &  $C(T) $ & $a_{in} $  & $a_{w}$ & $C_{max} $ \\
          \hline
             0  &  0 &  $1$ &  $1$ &  $20$ \\	
	    \hline
    \end{tabular}
    \small
    \caption{Fast Storage Specification.}\label{tab:storage:spec}
\end{table}

%% file: Tables/Storage.tex
\begin{table}[ht!]
    \centering\footnotesize
		\resizebox{\textwidth}{!}{
        \begin{tabular}{*{10}{|c|ccc|ccc|ccc}}
					\hline	
										\multicolumn{1}{|c}{} & \multicolumn{3}{|c}{Algorithm \ref{alg:cs}} & \multicolumn{3}{|c}{Algorithm \ref{alg:ct}} & \multicolumn{3}{|c|}{Algorithm \ref{alg:qdz}}\\
					\hline
$N$ & $\text{Price} $ & $\text{LSMC}$ & $\text{PATH}$ & $\text{Price} $ & $\text{LSMC}$ & $\text{PATH}$ & $\text{Price} $ & $\text{LSMC}$ & $\text{PATH}$\\
					\hline
$10^3$ & $282.7$ & $8.5$ & $0.3$ & $282.9$ & $15.3$ & $6.03$ & $281.3$ & $16.7$ & $7.4$\\
$10^4$ & $285.6$ & $136.3$ & $1.8$ & $286.7$ & $186.6$ & $57.09$ & $285.8$ & $198.7$ & $69.2$\\
$2\times 10^4$ & $287.7$ & $237.2$ & $3.3$ & $287.2$ & $357.1$ & $113.85$ & $287.0$ & $381.3$ & $138.1$\\
$5\times 10^4$ & $287.4$ & $674.7$ & $8.3$ & $286.8$ & $1002.1$ & $290.34$ & $286.8$ & $1063.7$ & $351.9$\\
$10^5$ & $287.5$ & $1197.0$ & $17.9$ & $287.0$ & $1818.3$ & $582.05$ & $286.9$ & $1936.5$ & $700.3$\\
					\hline
        \end{tabular}				
		}
    \scriptsize
    \caption{\footnotesize{Storages: Overall CPU times in seconds (LSMC), CPU times relative to the path generation (PATH) and comparison among the prices obtained with the different algorithms}}\label{tab:Storage:MC}
\end{table}


%% file: Tables/TableGasSwingSpec.tex
\begin{table}
    \centering
    \begin{tabular}{|c|c|c|c|c|}
            \hline
             $ACQmin$   &  $ACQmax$ & $DCQmin$  & $DCQmax$ & $T (days)$ \\
          \hline
             120  &  120 &  $1$ &  $1$ &  $360$ \\	
	    \hline
    \end{tabular}
    \small
    \caption{Specification of 120-120 take-or-pay Swing}\label{tab:swing:spec}
\end{table}

%% file: Tables/Swing.tex
\begin{table}[ht!]
    \centering\footnotesize
		\resizebox{\textwidth}{!}{
        \begin{tabular}{*{10}{|c|ccc|ccc|ccc}}
					\hline	
										\multicolumn{1}{|c}{} & \multicolumn{3}{|c}{Algorithm \ref{alg:cs:laplace}} & \multicolumn{3}{|c}{Algorithm \ref{alg:ct:laplace}} & \multicolumn{3}{|c|}{Algorithm \ref{alg:symm:bgou:qdz}}\\
					\hline
$N$ & $\text{Price} $ & $\text{LSMC}$ & $\text{PATH}$ & $\text{Price} $ & $\text{LSMC}$ & $\text{PATH}$ & $\text{Price} $ & $\text{LSMC}$ & $\text{PATH}$\\
					\hline
$10^3$ & $118.0$ & $7.2$ & $0.2$ & $117.2$ & $12.3$ & $5.32$ & $120.2$ & $13.6$ & $5.3$\\
$10^4$ & $117.6$ & $137.2$ & $1.2$ & $117.1$ & $181.8$ & $50.45$ & $118.7$ & $193.9$ & $50.4$\\
$2\times 10^4$ & $117.9$ & $237.4$ & $2.4$ & $117.5$ & $335.9$ & $102.04$ & $118.4$ & $360.1$ & $102.0$\\
$5\times 10^4$ & $118.0$ & $625.4$ & $5.7$ & $117.9$ & $849.2$ & $257.82$ & $118.5$ & $910.8$ & $257.8$\\
$10^5$ & $118.0$ & $1147.2$ & $11.7$ & $117.6$ & $1603.5$ & $519.85$ & $118.2$ & $1721.7$ & $519.8$\\
					\hline
        \end{tabular}				
		}
    \scriptsize
    \caption{\footnotesize{Swings: Overall CPU times in seconds (LSMC), CPU times relative to the path generation (PATH) and comparison among the prices obtained with the different algorithms}}\label{tab:Swing:MC}
\end{table}


%% file: Cufaro_Sabino_Fast_Pricing.bbl
\begin{thebibliography}{10}

\bibitem{BBP07}
O.~Bardou, S.~Bouthemy, and G.~Pag\'es.
\newblock Optimal {Q}uantization for the {P}ricing of {S}wing {O}ptions.
\newblock {\em Applied Mathematical Finance}, 16(2):183--217, 2009.

\bibitem{BNSh01}
O.E. Barndorff-Nielsen and N.~Shephard.
\newblock Non-{G}aussian {O}rnstein-{U}hlenbeck-based {M}odels and {S}ome of
  {T}heir {U}ses in {F}inancial {E}conomics.
\newblock {\em Journal of the Royal Statistical Society: Series B},
  63(2):167--241, 2001.

\bibitem{BBKL2007}
H.~Ben-Ameur, M.~Breton, L.~Karoui, and P.~L'Ecuyer.
\newblock {A {D}ynamic {P}rogramming {A}pproach for {P}ricing {O}ptions
  {E}mbedded in {B}onds}.
\newblock {\em Journal of Economic Dynamics and Control}, 31(7):2212--2233,
  July 2007.

\bibitem{BenthPircalabu18}
F.E. Benth and A.~Pircalabu.
\newblock A non-gaussian {O}rnstein-{U}hlenbeck {M}odel for {P}ricing {W}ind
  {P}ower {F}utures.
\newblock {\em Applied Mathematical Finance}, 25(1), 2018.

\bibitem{BMBK07}
F.E. Benth and J.~Kallsen T.~Meyer-Brandis.
\newblock A non-{G}aussian {O}rnstein-{U}hlenbeck {P}rocess for {E}lectricity
  {S}pot {P}rice {M}odeling and {D}erivatives {P}ricing.
\newblock {\em Applied Mathematical Finance}, 14(2):153--169, 2007.

\bibitem{Bertsekas05}
D.~P. Bertsekas.
\newblock {\em Dynamic {P}rogramming and {O}ptimal {C}ontrol, {V}olume {I}}.
\newblock Athena Scientific, Belmont, Mass., third edition, 2005.

\bibitem{BDJ08}
A.~Boogert and C.~de~Jong.
\newblock Gas {S}torage {V}aluation {U}sing a {M}onte {C}arlo {M}ethod.
\newblock {\em Journal of Derivatives}, 15:81--91, 2008.

\bibitem{CarteaFigueroa}
A.~Cartea and M.~Figueroa.
\newblock Pricing in {E}lectricity {M}arkets: a {M}ean {R}everting {J}ump
  {D}iffusion {M}odel with {S}easonality.
\newblock {\em Applied Mathematical Finance, No. 4, December 2005},
  12(4):313--335, 2005.

\bibitem{ContTankov2004}
R.~Cont and P.~Tankov.
\newblock {\em Financial {M}odelling with {J}ump {P}rocesses}.
\newblock Chapman and Hall, 2004.

\bibitem{cufaro08}
N.~{Cufaro Petroni}.
\newblock Self-decomposability and {S}elf-similarity: a {C}oncise {P}rimer.
\newblock {\em Physica A, Statistical Mechanics and its Applications},
  387(7-9):1875--1894, 2008.

\bibitem{cs17}
N.~{Cufaro Petroni} and P.~Sabino.
\newblock Coupling {P}oisson {P}rocesses by {S}elf-decomposability.
\newblock {\em Mediterranean Journal of Mathematics}, 14(2):69, 2017.

\bibitem{CufaroSabino:QF18}
N.~{Cufaro Petroni} and P.~Sabino.
\newblock Pricing exchange options with correlated jump diffusion processes.
\newblock {\em Quantitative Finance}, pages 1--13, 2018.

\bibitem{cs20}
N.~{Cufaro Petroni} and P.~Sabino.
\newblock Gamma {R}elated {O}rnstein–{U}hlenbeck {P}rocesses and their
  {S}imulation.
\newblock available at: https://arxiv.org/abs/2003.08810, 2020.

\bibitem{CKM17}
M.~Cummins, G.~Kiely, and B.~Murphy.
\newblock Gas {S}torage {V}aluation under {L}\'{e}vy {P}rocesses using {F}ast
  {F}ourier {T}ransform.
\newblock {\em Journal of Energy Markets}, 4:43--86, 2017.

\bibitem{Deng00stochasticmodels}
S.~Deng.
\newblock Stochastic {M}odels of {E}nergy {C}ommodity {P}rices and {T}heir
  {A}pplications: {M}ean-reversion with {J}umps and {S}pikes.
\newblock Citeseer, 2000.

\bibitem{HHM11}
B.~Hambly, S.~Howison, and T.~Kluge.
\newblock Information-{B}ased {M}odels for {F}inance and {I}nsurance.
\newblock {\em Quantitative Finance}, 9(8):937--949, 2009.

\bibitem{Kjaer2008}
M.~Kjaer.
\newblock Pricing of {S}wing {O}ptions in a {M}ean {R}everting {M}odel with
  {J}umps.
\newblock {\em Applied Mathematical Finance}, 15(5-6):479--502, 2008.

\bibitem{Kluge2006}
T.~Kluge.
\newblock {P}ricing {S}wing {O}ptions and other {E}lectricity {D}erivatives.
\newblock Technical report, University of Oxford, 2006.
\newblock PhD Thesis, Available at
  http://perso-math.univ-mlv.fr/users/bally.vlad/publications.html.

\bibitem{Kou2002}
S.~G. Kou.
\newblock A {J}ump-{D}iffusion {M}odel for {O}ption {P}ricing.
\newblock {\em Manage. Sci.}, 48(8):1086--1101, August 2002.

\bibitem{KT2008}
U.~K\"{u}chler and S.~Tappe.
\newblock Bilateral {G}amma {D}istributions and {P}rocesses in {F}inancial
  {M}athematics.
\newblock {\em Stochastic Processes and their Applications}, 118(2):261--283,
  2008.

\bibitem{LindskogMcNeil}
F.\ Lindskog and J.\ McNeil.
\newblock Common poisson shock models: applications to insurance and credit
  risk modelling.
\newblock {\em ASTIN Bulletin}, 33(2):209--238, 2003.

\bibitem{LSW01}
F.~A. Longstaff and E.S. Schwartz.
\newblock Valuing {A}merican {O}ptions by {S}imulation: a {S}imple
  {L}east-{S}quares {A}pproach.
\newblock {\em Review of Financial Studies}, 14(1):113--147, 2001.

\bibitem{LS02}
J.J. Lucia and E.S. Schwartz.
\newblock Electricity {P}rices and {P}ower {D}erivatives: {E}vidence from the
  {N}ordic {P}ower {E}xchange.
\newblock {\em Review of Derivatives Research}, 5(1):5--50, Jan 2002.

\bibitem{McK87}
E.~McKenzie.
\newblock Innovation {D}istridution for {G}amma and {N}egative {B}inomial
  {A}utoregressions.
\newblock {\em Scandinavian Journal of Statistics: Theory and Applications},
  14(1):79--85, 1987.

\bibitem{MBT2008}
T.~Meyer-Brandis and P.~Tankov.
\newblock Multi-factor {J}ump-diffusion {M}odels of {E}lectricity {P}rices.
\newblock {\em International Journal of Theoretical and Applied Finance},
  11(5):503--528, 2008.

\bibitem{PellegrinoSabino15}
T.~Pellegrino and P.~Sabino.
\newblock Enhancing {L}east {S}quares {M}onte {C}arlo with {D}iffusion
  {B}ridges: an {A}pplication to {E}nergy {F}acilities.
\newblock {\em Quantitative Finance}, 15(5):761--772, 2015.

\bibitem{QDZ19}
Y.~Qu, A.~Dassios, and H.~Zhao.
\newblock Exact {S}imulation of {G}amma-driven {O}rnstein–{U}hlenbeck
  {P}rocesses with {F}inite and {I}nfinite {A}ctivity {J}umps.
\newblock {\em Journal of the Operational Research Society}, 0(0):1--14, 2019.

\bibitem{Sabino20}
P.~Sabino.
\newblock Forward or {B}ackward {S}imulations? {A} {C}omparative {S}tudy.
\newblock {\em Quantitative Finance}, 2020.
\newblock In press.

\bibitem{Sato}
K.\ Sato.
\newblock {\em L\'evy {P}rocesses and {I}nfinitely {D}ivisible
  {D}istributions}.
\newblock Cambridge U.P., Cambridge, 1999.

\bibitem{Schoutens03}
W.~Schoutens.
\newblock {\em L\'{e}vy Processes in Finance: Pricing Financial Derivatives}.
\newblock John Wiley and Sons Inc, 2003.

\bibitem{SchwSchm00}
P.~Schwartz and J.E. Smith.
\newblock Short-term {V}ariations and {L}ong-term {D}ynamics in {C}ommodity
  {P}rices.
\newblock {\em Management Science}, 46(7):893--911, 2000.

\end{thebibliography}
